# Structure-Property-Performance Relationships of Dielectric Nanostructures for Mie Resonance-Enhanced Dye-Sensitization

*Ravi Teja Addanki Tirumala*[⊥]*, Sundaram Bhardwaj Ramakrishnan*[⊥]*, Farshid Mohammadparast*[⊥]*, Swetha M. Arumugam*[§]*, Susheng Tan*[†]*, Marimuthu Andiappan\**[⊥]

Affiliations**:

[⊥] School of Chemical Engineering, Oklahoma State University, Stillwater, OK, USA.

[§] Department of Chemistry, PSG College of Arts & Science, Coimbatore, Tamil Nadu, India.

[†] Department of Electrical and Computer Engineering and Petersen Institute of Nano Science and Engineering, University of Pittsburgh, Pittsburgh, PA, USA

\* **Corresponding Author**, Email: mari.andiappan@okstate.edu

## ABSTRACT

Dye-sensitized photocatalytic (DSP) approach is considered as one of the promising approaches for developing visible light- and near-infrared light-responsive photocatalysts. DSP systems are still affected by significant drawbacks, such as low light absorption efficiency. Recently, it has been demonstrated that the plasmonic metal nanostructures can be used to enhance the light absorption efficiency and the overall dye-sensitization rate of DSP systems through the plasmonic Mie resonance-enhanced dye-sensitization approach. In this contribution, we report an alternate and novel approach, dielectric Mie resonance-enhanced dye sensitization. Specifically, we demonstrate that the dielectric Mie resonances in cuprous oxide ($Cu_2O$) spherical and cubical nanostructures can be used to enhance the dye-sensitization rate of methylene blue dye. The $Cu_2O$ nanostructures exhibiting dielectric Mie resonances exhibit up to an order of magnitude higher dye-sensitization rate as compared to $Cu_2O$ nanostructures not exhibiting dielectric Mie resonances. Our model system developed from finite-difference time-domain simulation predicts a volcano-type relationship between the dye sensitization rate and the size of $Cu_2O$ nanostructures. The predicted structure-property-performance relationship is experimentally verified and the optimal size ranges of $Cu_2O$ nanospheres and nanocubes are identified. Although we demonstrate the dielectric Mie resonance-enhanced dye-sensitization approach using $Cu_2O$ nanostructures, the proposed approach can be used to design a wide range of DSP systems, including $CeO_2$, $\alpha$-$Fe_2O_3$, and $TiO_2$ nanostructures-based DSP systems.

**GRAPHICAL TABLE OF CONTENTS (TOC Graphic)**

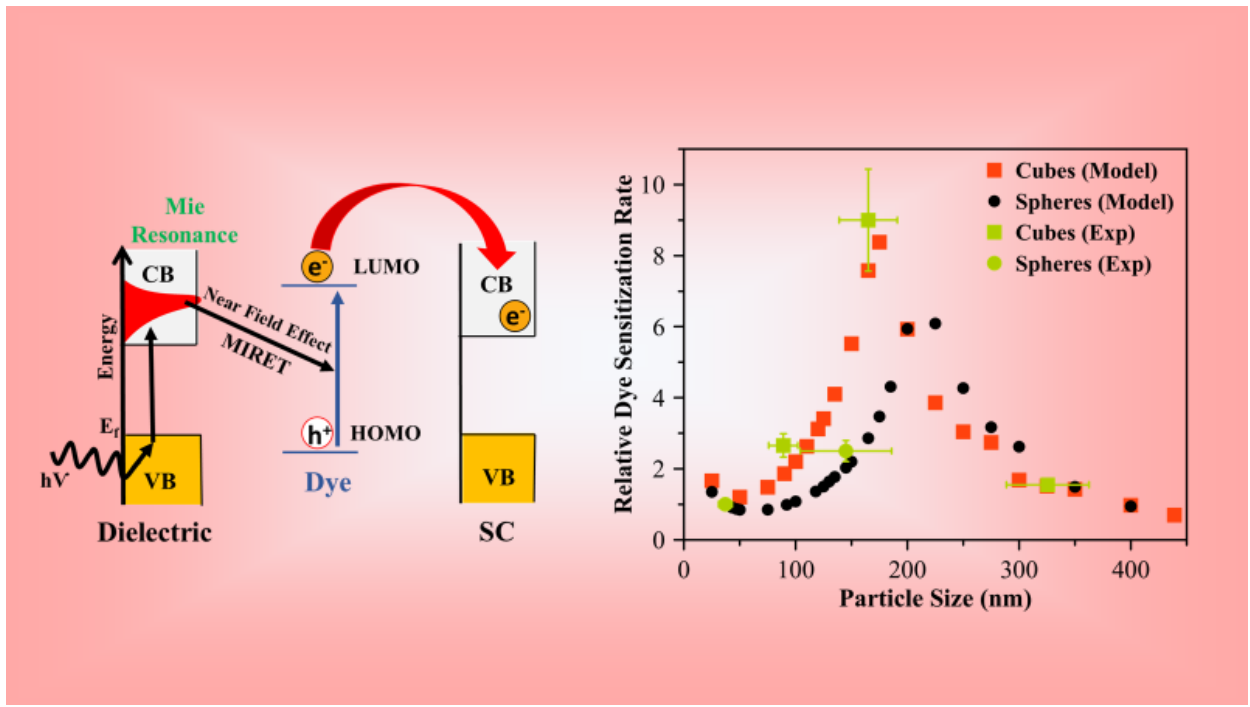

## MAIN TEXT

Dye-sensitized photocatalytic (DSP) systems have emerged as a promising approach for a number of applications, including renewable hydrogen ($H_2$) production via photocatalytic water splitting, photocatalytic conversion of carbon dioxide ($CO_2$) into value-added products, and pollution mitigation. In a typical DSP system, a dye molecule is used as a photosensitizer that is adsorbed on a semiconductor nanostructure. As shown in Figure 1a, in the DSP system, the dye molecules absorb incident light and inject the excited electrons into the conduction band (CB) of the semiconductor. These excited electrons can be used to drive reduction reactions, such as the reduction of water for $H_2$ production. The regeneration of the dye molecule for cyclic utilization can be accomplished using the electron donor. In the last two decades, significant progress has been made in the field, and studies have shown that by appropriately selecting and matching the energy levels of dye molecules and semiconductors, DSP systems can be designed to cover the visible region and even the near-infrared region of the solar spectrum. Despite these great successes achieved in the field, DSP systems are still affected by significant drawbacks, such as low light absorption efficiency.[1–5]

Recently, it has been demonstrated that plasmonic Mie resonances of metal nanostructures, such as silver (Ag) and gold (Au) can be used to enhance the light absorption efficiency of dye molecules in the dye-sensitized photocatalytic and photovoltaic systems.[6–17] The plasmonic metal nanostructures (PMNs) under Mie resonance conditions can exhibit very high absorption cross section values that are up to 5 orders of magnitude higher than dye molecules.[18–20] Therefore, PMN (e.g., Ag and Au) can harvest a large fraction of the incident light and transfer the energy into the nearby dye molecules and enhance their light absorption efficiency via a number of pathways, including nanoantenna effect and plasmon-induced resonance energy transfer (Figure 1b).[21–23]

These plasmonic Mie resonance-mediated effects are utilized to enhance the rate of dye sensitization in DSP systems. However, these PMNs-based DSP systems also possess inherent challenges, such as compatibility issues with conventional semiconductor manufacturing, high material costs, and issues of commercialization with increasing complexity due to the need for both metals and semiconductors.

Herein, we propose a solution to the aforementioned issues through an alternate approach, dielectric Mie resonance-enhanced dye sensitization shown schematically in Figure 1c. The Mie resonances can occur in plasmonic materials that exhibit negative permittivity ($\varepsilon < 0$) as well as dielectric ($\varepsilon > 0$) materials with moderate (2.5-3.5) and high refractive index values (>3.5).[24–32] The dielectric Mie resonance-enhanced dye-sensitization proposed in Figure 1c is, therefore, most suitable for DSP systems built on metal oxide semiconductors with moderate and high refractive index values. Such metal oxide semiconductors include cerium(IV) oxide ($CeO_2$), cuprous oxide ($Cu_2O$), hematite iron oxide ($\alpha\text{-}Fe_2O_3$), and titanium dioxide ($TiO_2$).

Similar to the plasmonic Mie resonances of PMNs, the dielectric Mie resonances of nanostructures of medium- and high-refractive-indexed metal oxide semiconductors can transfer the photonic energy into the adsorbed dye molecules via pathways such as nanoantenna effect, resonance energy transfer, and Mie resonance-mediated intense scattering effect. These dielectric Mie-resonance effects are expected to enhance the light absorption efficiency of the dye molecules in the DSP systems (Figure 1c). One major difference is that, unlike the plasmonic Mie resonance-based system (Figure 1b), in a dielectric Mie resonance-based system (Figure 1c), there is no need for a separate light-enhancing material. Being a major advantage, this system solely requires a dielectric semiconductor nanostructure and dye. Specifically, the dielectric semiconductor nanostructure can play the role of a Mie resonator for enhancing the light absorption efficiency of the dye molecule as well as a source for harvesting excited electrons from the dye molecules and facilitating the reduction reaction (Figure 1c).

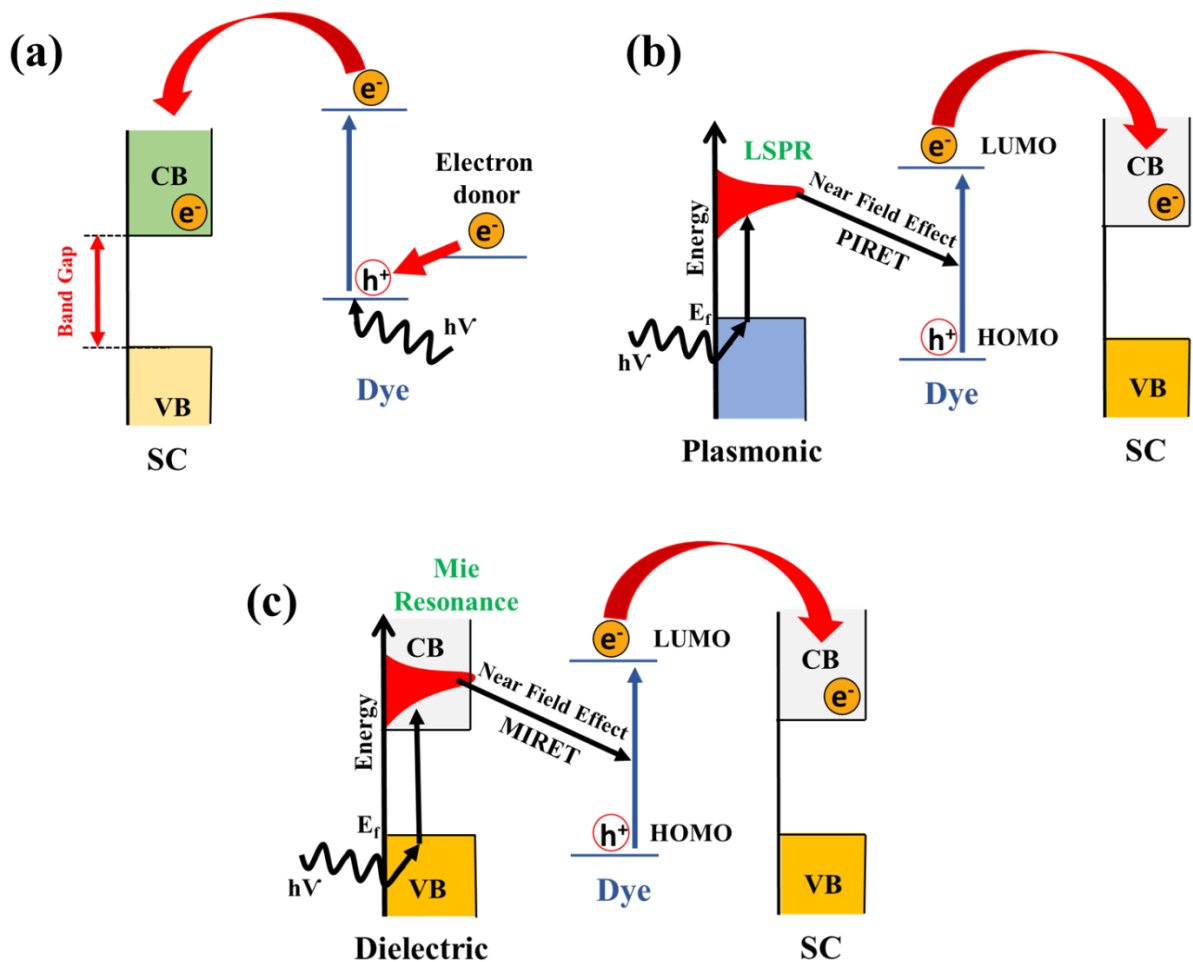

**Figure 1.** Schematic diagram of **(a)** dye-Sensitization and electron transfer into the conduction band (CB) of semiconductor (SC), **(b)** localized surface plasmon resonance (LSPR)-enhanced dye-sensitization, and (**c**) dielectric Mie resonance-enhanced dye-sensitization.

In this contribution, we demonstrate the dielectric Mie resonance-enhanced dye-sensitization approach using methylene blue (MB) dye-sensitization on $Cu_2O$ nanostructures as an example. We have developed the structure-property-performance relationships of $Cu_2O$ nanostructures for MB dye-sensitization. Specifically, different sizes of $Cu_2O$ nanospheres and nanocubes are synthesized and the structure of these $Cu_2O$ nanostructures are characterized using transmission electron microscopy (TEM) and scanning electron microscopy (SEM). The size-dependent optical properties of the $Cu_2O$ nanospheres and nanocubes are predicted using finite-difference time-domain (FDTD) simulations. Based on these FDTD simulation results, we predict a volcano-type relationship between the size of $Cu_2O$ nanostructures and the enhancement of dye-sensitization rate by the respective nanostructures. The predicted structure-property-performance relationships of $Cu_2O$ nanospheres and nanocubes are experimentally verified by measuring the rate of MB dye-sensitization followed by dye-degradation (DSD).

The procedures followed for FDTD simulations, syntheses and characterizations of $Cu_2O$ nanospheres and nanocubes,[24,33,34] and MB dye-sensitization and degradation experiments are detailed in the supporting information (SI). Briefly, $Cu_2O$ nanospheres with average diameters of 37 nm and 145 nm and $Cu_2O$ nanocubes with average edge lengths of 92 nm, 165 nm and 325 nm were synthesized using the synthesis protocols reported in our previous contributions.[24,33,34] The sizes and $Cu_2O$ phases of the nanostructures were confirmed using TEM and SEM, and X-ray diffraction patterns, respectively (Figure S1a-b in SI). The performance of the $Cu_2O$ nanostructures towards MB dye-sensitization was evaluated through the measurement of the rate of MB degradation that occurs in the MB dye-sensitization region. The MB degradation was carried out in the solution phase using dimethylformamide (DMF) as a solvent. To quantify the extent of MB degradation, the concentration (C) of MB in the reaction mixture was quantified as a function of irradiation time. The MB absorption value at its peak absorption wavelength (i.e., 665 nm) in the ultraviolet-visible (UV-Vis) absorption spectrum was used to quantify the concentration of MB.

Figure 2a-d shows the representative TEM and SEM images of quasi-spherical $Cu_2O$ nanoparticles with an average diameter of 37 nm and 145 nm, and $Cu_2O$ nanocubes with an average edge length

of 165 and 325 nm, respectively. Figure 2e-f shows the experimentally measured UV-Vis-near IR extinction spectra of these $Cu_2O$ nanostructures. The representative SEM image and UV-Vis-near IR extinction spectra of $Cu_2O$ nanocubes of 92 nm average edge length are also provided in Figure S1d-S1e in SI. The extinction spectra shown in Figures 2e-f are consistent with the extinction features predicted from the FDTD simulations (Figure S2a-d in SI). Specifically, as seen from Figure 2e, $Cu_2O$ nanospheres of 37 nm average diameter exhibit extinction features similar to the bulk $Cu_2O$, which is a semiconductor with a bandgap of 2.1 eV.[24] For these 37 nm $Cu_2O$ nanospheres, the dielectric Mie resonances are absent in the UV-Vis-near IR regions. In contrast, $Cu_2O$ nanospheres and nanocubes of sizes larger than 90 nm exhibit strong dielectric Mie resonances in the UV-Vis-near IR regions, as shown in Figure 2e-f (also see Figure S1c and Figure S2a-d in SI). The lowest Mie resonance peak in the UV-Vis-near IR extinction spectrum corresponds to the combination of the electric dipole and magnetic dipole (Figure S2e in SI).[24] Similarly, the second-lowest energy peak and higher-order peaks correspond to the combination of electric quadrupole and magnetic quadrupole, and the combination of higher-order electric and magnetic modes, respectively (Figure S2f in SI).[24]

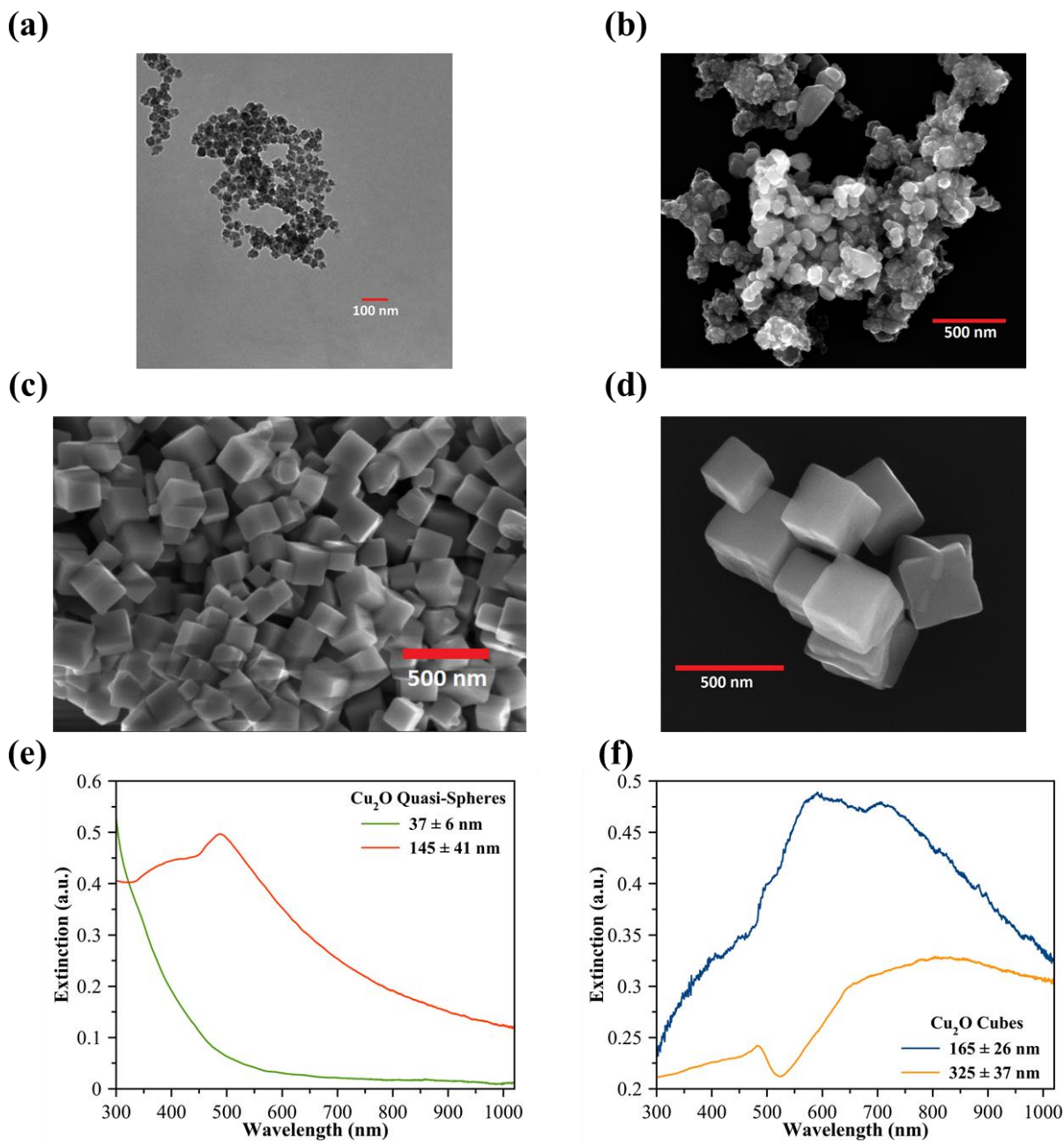

**Figure 2.** (**a-d**) Representative TEM/SEM images of (**a**) quasi-spherical $Cu_2O$ nanoparticles of $37 \pm 6$ nm diameter, (**b**) quasi-spherical $Cu_2O$ nanoparticles of $145 \pm 41$ nm diameter, (**c**) $Cu_2O$ nanocubes of $165 \pm 26$ nm edge length, and (**d**) $Cu_2O$ nanocubes of $325 \pm 37$ nm edge length. (**e-f**) UV-Vis-near IR extinction spectra of (**e**) quasi-spherical $Cu_2O$ nanoparticles of $37 \pm 6$ nm and 145 ± 41 nm diameters, and (**f**) $Cu_2O$ nanocubes of 165 ± 26 nm and 325 ± 37 nm edge lengths, dispersed in DMF.

To evaluate the performance of the $Cu_2O$ nanostructures towards the MB DSD, we have selected red LEDs as the illumination source. The intensity of the light, when measured at the surface of the photoreactor, was 7.49 mW/cm$^2$, and it was kept constant for all dye-sensitization experiments reported in this contribution. In Figure 3a, we show the spectrum of red LEDs and UV-Vis absorption spectrum of MB dye. As seen from Figure 3a, there is a significant overlap between the spectrum of red LEDs and the absorption spectra of methylene blue in the 590-670 nm region. It has been previously shown that, in this region (590-670 nm), MB molecules can undergo dye-sensitization and inject the excited electrons into the conduction band of the semiconductor (Figure 3b)[35,36]. In the presence of dissolved oxygen ($O_2$), these electrons can form superoxide ($O_2^-$), which can attack and cause the degradation of excited MB molecules (Figure 3b) [35,36]. Since the bandgap of $Cu_2O$ is 2.1 eV (~590 nm)[24,37], the rate of MB degradation via direct photocatalysis by $Cu_2O$ nanostructures is expected to be minimal in this MB dye-sensitization region (590-670 nm) investigated in this study.

To illustrate the role of dielectric Mie resonances on the MB dye-sensitization, we carried out the experiments to quantify the rate of MB DSD using quasi-spherical $Cu_2O$ nanoparticles with an average diameter of 145 nm and $Cu_2O$ nanocubes with an average edge length of 92 nm, 165 nm, and 325 nm. These $Cu_2O$ nanostructures exhibit dielectric Mie resonances in the MB dye-sensitization region (590-670 nm), as seen from Figure 2e-f. We also carried out the same experiments using quasi-spherical $Cu_2O$ nanoparticles with an average diameter of 37 nm, in which the dielectric Mie resonances are absent in the MB dye sensitization region (Figure 2e). In these dye-sensitization experiments performed with different sizes of $Cu_2O$ nanospheres and nanocubes, the weight load of the $Cu_2O$ nanostructures in the reaction mixture was kept constant. Before exposing the reaction mixture to the red-light illumination, the MB dye molecules are stirred with $Cu_2O$ nanostructures dispersed in DMF for 3 hours to reach the adsorption equilibrium. Also, the reaction mixture is sparged with air for 30 minutes to start with the same level of dissolved oxygen for all experiments. Figure 3c shows the extent of MB degradation ($C/C_0$) under the red-light illumination (590-670 nm) conditions in the presence of different sizes of $Cu_2O$ nanospheres and nanocubes, and also in the absence of $Cu_2O$ nanostructures.

As seen from Figure 3c, the slowest rate of MB degradation is observed for the dye-only conditions carried in the absence of $Cu_2O$ nanostructures. For the experiments performed in the presence of different sizes of $Cu_2O$ nanospheres and nanocubes, the $Cu_2O$ nanostructures with dielectric Mie resonances exhibit a higher rate of MB degradation than the 37 nm quasi-spherical $Cu_2O$ nanoparticles that exhibit no Mie resonance in the dye sensitization region. Specifically, the increasing rate of MB degradation is observed in the following order: 165 nm $Cu_2O$ nanocubes > 92 nm $Cu_2O$ nanocubes > 145 nm quasi-spherical $Cu_2O$ nanoparticles > 325 nm $Cu_2O$ nanocubes > 37 nm quasi-spherical $Cu_2O$ nanoparticles (also see Table S1 in SI). To investigate the possible role of the light-induced heating effect on MB degradation, we measured the light-induced increase in the temperature of the reaction mixture. We found that the temperature of the reaction mixture was increased from ~20 °C to ~30 °C under MB dye-sensitization conditions (Figure S3a-e in SI). This increase in the temperature was uniform for all experiments performed in the presence of different sizes of $Cu_2O$ nanospheres and nanocubes as well as for the experiments performed in the absence of $Cu_2O$ nanostructures (Figure S3a-e in SI). We also performed heating experiments at the elevated temperature of 60 °C. No significant degradation was observed in these heating experiments performed under dark conditions in the absence of red-light irradiation (Figure S3f in SI). These results confirm that the difference in the rate of MB degradation observed in the presence of Cu₂O nanospheres and nanocubes of different sizes is not due to the light-induced heating effect. Therefore, we attribute the higher rate of MB degradation observed on the Cu₂O nanostructures with dielectric Mie resonances (e.g., 165 nm Cu₂O nanocubes) in Figure 3c to the dielectric Mie resonance-enhanced MB dye-sensitization.

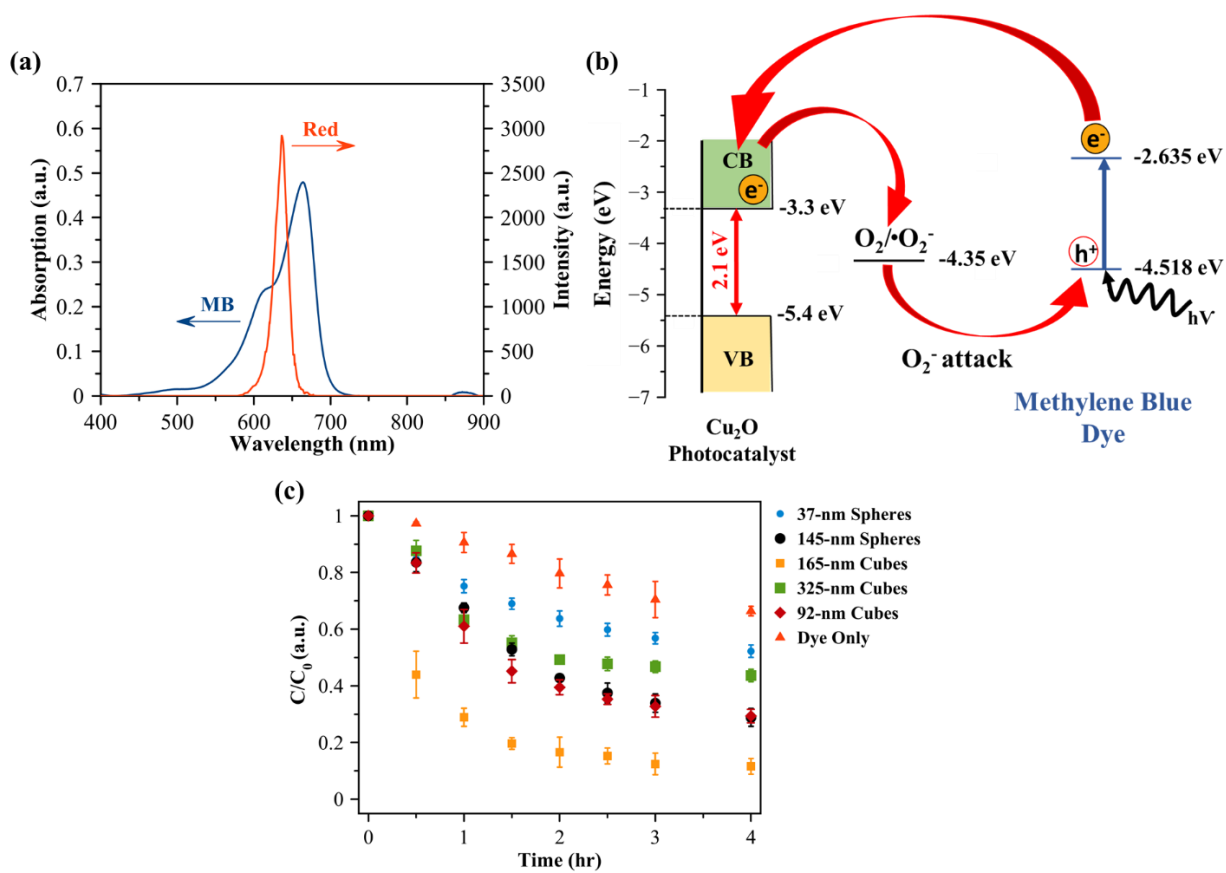

**Figure 3.** **(a)** Absorption spectrum of MB (left ordinate) and spectrum of red LED light source used for MB dye-sensitization (right ordinate). **(b)** Schematic diagram illustrating the MB DSD via superoxide ($O_2^-$) intermediate species.[38,39] **(c)** Plot of $C/C_0$ versus irradiation time for MB DSD in DMF using quasi-spherical Cu₂O nanoparticles of $37 \pm 6$ nm diameter (blue circles), quasi-

spherical Cu₂O nanoparticles of 145 ± 41 nm diameter (black circles), Cu₂O nanocubes of 92 ± 13 nm edge length (maroon diamonds), Cu₂O nanocubes of 165 ± 26 nm edge length (orange squares), Cu₂O nanocubes of 325 ± 37 nm edge length (green squares), and under blank conditions in the absence of photocatalyst (red triangles).

For the dye-only conditions, in the absence of Cu₂O nanostructures, the MB degradation can occur via the absorption of the incident photons by the dye molecules followed by the photoexcited electron-transfer from the excited MB molecules into the dissolved oxygen to form superoxide radicals ($O_2^-$). The superoxide radicals can then attack and degrade the excited MB molecules (Figure S4a in SI).[35] The Cu₂O nanostructures can enhance the light absorption efficiency of the dye molecules via a number of pathways, including the electromagnetic nearfield or nanoantenna effect and Mie resonance-induced resonance energy transfer (MIRET). The Cu₂O nanostructures can also facilitate and enhance the electron transfer between adsorbed MB and O₂ molecules and the subsequent degradation of the excited MB molecules. In this contribution, to develop the structure-property-performance relationships of Cu₂O nanostructures for MB DSD, we propose the following simple approximation as shown in equation (1) below (see SI for more details). This equation relates the rate of MB dye degradation that occurs vis dye-sensitization pathway in the presence of Cu₂O nanostructures and the optical and geometrical properties of the respective Cu₂O nanostructures.

$$ r \propto \int I_0(\lambda) Abs_{MB}(\lambda) E_{Cu2O}(\lambda) \mathrm{d}\lambda \times \frac{S}{V} \qquad (1) $$

where, r is the rate of dye-degradation per unit mass of the Cu₂O nanostructures, Ext$_{Cu2O}$ is the volume-normalized extinction cross section of the Cu₂O nanostructure, Abs$_{MB}$ is the absorbance of MB, $I_0$ is the intensity of the incident light, and $\lambda$ is the wavelength of the incident light, and S and V are the surface area and volume of the $Cu_2O$ nanostructure.

To predict the structure-property-performance relationships of $Cu_2O$ nanostructures using equation (1), we performed FDTD simulations of $Cu_2O$ nanospheres and nanocubes of different sizes (25-400 nm) and simulated the extinction cross section of the respective nanostructures as a function of incident light wavelength. The representative FDTD-simulated extinction spectra of these $Cu_2O$ nanocubes are provided in Figure 4a (also see Figure S4b-e in SI). Based on equation (1), in Figure 4a, the extinction cross section values are normalized to the fourth power of the edge length. These FDTD-simulated values are used in equation (1) to predict the performance of $Cu_2O$ nanospheres and nanocubes of different sizes in the range of 25 to 400 nm. Figure 4b shows the rate of MB degradation that can occur in the dye-sensitization region (590-670 nm) in the presence of these $Cu_2O$ nanostructures. In Figure 4b, the rate values in the y-axis are normalized with respect to the $Cu_2O$ nanosphere of 37 nm diameter. As seen from Figure 4b, our model that is based on equation (1) and FDTD-simulated extinction cross section values predicts a volcano-type relationship between the rate of DSD and the size of $Cu_2O$ nanostructures. From the kinetic data collected from the experiments performed with quasi-spherical $Cu_2O$ nanoparticles of 37 nm and 145 nm average diameters and $Cu_2O$ nanocubes of 92 nm, 165 nm and 325 nm edge lengths (Figure 3c), we calculated the relative rate of DSD for these $Cu_2O$ nanostructures (see Figure S4v-z and Table 1 in SI for more details). As seen from Figure 4b, the experimentally measured values are in very well agreement with the predicted values. The optimum sizes of $Cu_2O$ nanospheres and nanocubes in the volcano plot shown in Figure 4b are in the range of 165-200 nm. The main reason for these $Cu_2O$ nanostructures to appear in the optimum range is their high extinction cross section values in the MB dye sensitization region (i.e., 590-670 nm), as seen from Figure 4a. Moreover, the dielectric Mie resonance-enhanced dye-sensitization causes the 165-nm nanocubes to exhibit approximately an order of magnitude (~9 times) higher dye-sensitization rate as compared to 37-nm nanospheres, in which the Mie resonance is absent (see Figure 4b and Table S1 in SI).

The structure-property-performance relationships developed in this contribution can be used for designing a wide range of DSP systems. For example, when the $Cu_2O$ nanostructures are used for DSP systems consist of a dye that absorbs in the shorter wavelength region (e.g., 460-500 nm), our model shown in equation (1) predicts that optimum sizes of $Cu_2O$ nanostructures move towards smaller sizes (e.g., 120-150 nm $Cu_2O$ nanocubes in Figure 4a). Similarly, for dyes that can absorb in the near-IR wavelength region (e.g., 800-900 nm), the optimum sizes are predicted to move towards larger sizes (e.g., 240-270 nm $Cu_2O$ nanocubes in Figure 4a). The approach demonstrated in this study can also be used to design DSP systems that can involve a wide range of combinations of medium- and high-refractive-index semiconductors and dye molecules. Our FDTD simulation results shown in Figure S4f-u in SI predict that such combinations can include semiconductors such as $CeO_2$, $CuO$, $\alpha$-$Fe_2O_3$, and $TiO_2$ and appropriate dye molecules that can absorb anywhere in the visible and near-IR wavelength regions. The energy levels of these dye molecules need to be in alignment with the conduction band edge of the semiconductors so that the excited electrons can be transferred from the dye molecules into the semiconductor.

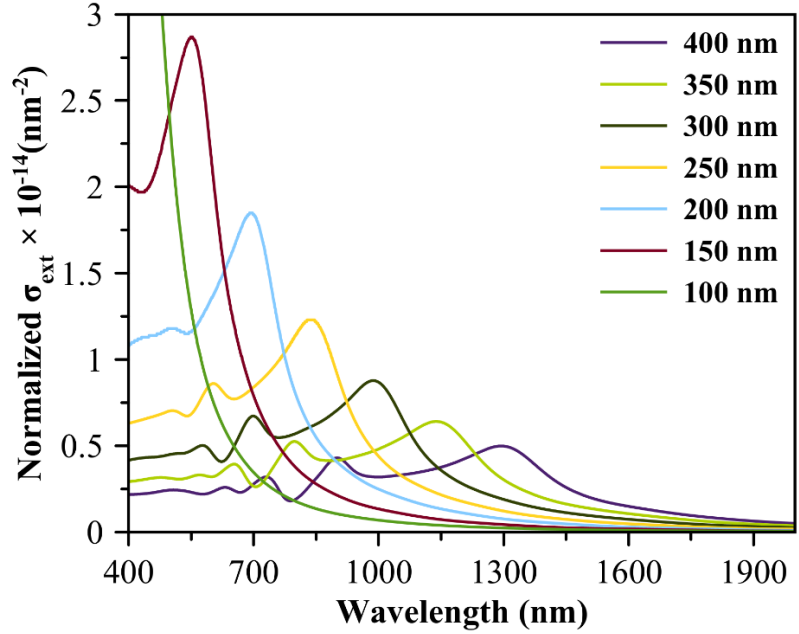

**Figure 4a**. FDTD-simulated normalized-extinction (normalized-$\sigma_{Ext}$) cross section of $Cu_2O$ nanocubes of different edge lengths in the range of 25 to 400 nm as a function of incident light wavelength. Normalized-$\sigma_{Ext}$ = Ratio of extinction cross section ($\sigma_{Ext}$) of nanocube to the fourth power of its edge length ($A^4$).

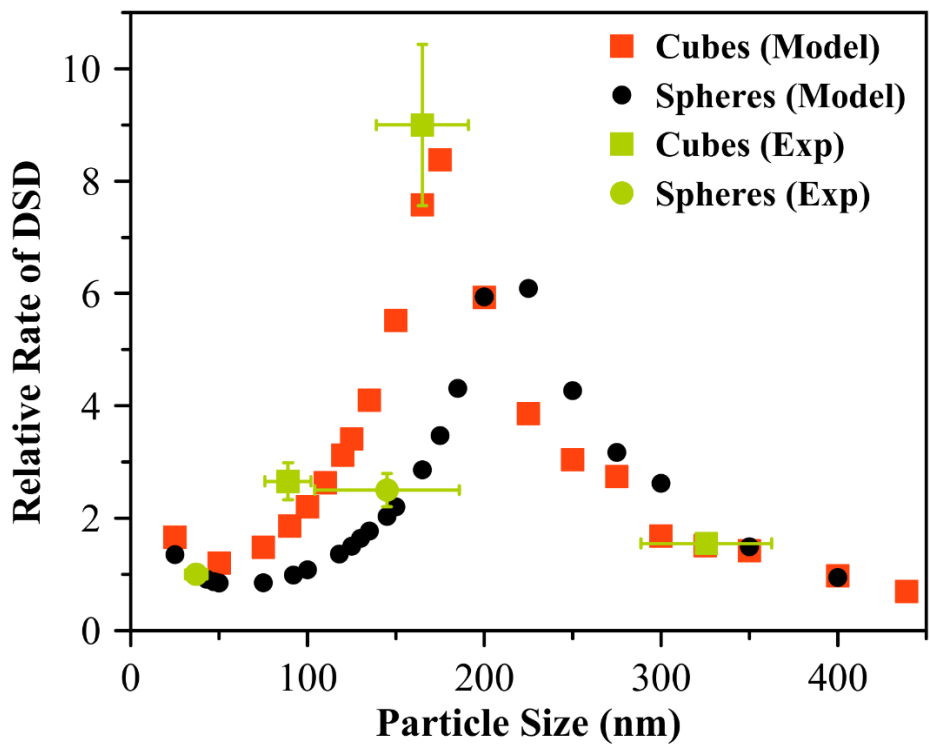

**Figure 4b.** Volcano plot showing the predicted and experimentally measured relative rate of MB DSD as a function of size of $Cu_2O$ nanospheres and nanocubes. Black circles and red squares represent predicted relative rates for $Cu_2O$ nanospheres and nanocubes, respectively. Green circles, and green squares represent experimentally measured relative rates on quasi-spherical $Cu_2O$ nanoparticles of $37 \pm 6$ nm diameter, quasi-spherical $Cu_2O$ nanoparticles of $145 \pm 41$ nm diameter, $Cu_2O$ nanocubes of $92 \pm 13$ nm edge length, $Cu_2O$ nanocubes of $165 \pm 26$ nm edge length and $Cu_2O$ nanocubes of $325 \pm 37$ nm edge length, respectively. Rate of MB DSD on quasi-spherical $Cu_2O$ nanoparticles of $37 \pm 6$ nm diameter is used as a reference.

In conclusion, we have demonstrated that the rate of dye-sensitization can be enhanced in the presence of semiconductor nanostructures with dielectric Mie resonances. Using FDTD simulation results, we have developed the structure-property-performance relationship for the dielectric Mie resonance-enhanced dye-sensitization. The predicted volcano-type relationship between the rate of MB dye-sensitization and the size of $Cu_2O$ nanostructures is experimentally verified using the rate measurements of MB DSD in the presence of $Cu_2O$ nanospheres and nanocubes of different sizes. The flexibility of tuning the dielectric Mie resonance peaks across the visible and near-IR regions by controlling the size of the dielectric nanostructures marks its applicability to a wide range of DSP systems. Specifically, the findings of this contribution open a novel approach to design efficient DSP systems that can involve semiconductors such as $CeO_2$, $CuO$, $Cu_2O$, $\alpha$-$Fe_2O_3$, and $TiO_2$ and a wide range of visible- and near-IR-responsive dye molecules.

## ASSOCIATED CONTENT

Syntheses and characterizations of $Cu_2O$ nanostructures, Experimental details of DSD experiments, Computational details for FDTD Simulations, and other details are listed as the contents of the supporting Information.

## AUTHOR INFORMATION

* Corresponding Author, Email: mari.andiappan@okstate.edu

**Notes**

The authors declare no competing financial interest.

## ACKNOWLEDGEMENTS

We gratefully acknowledge Dr. Andiappan's awarded funding from the National Science Foundation, CBET Catalysis Program under Grant No. NSF CBET-2102238. We also acknowledge funding from the Oklahoma Center for the Advancement of Science and Technology under the project number HR18-093.

# Structure-Property-Performance Relationships of Dielectric Nanostructures for Mie Resonance-Enhanced Dye-Sensitization

**Supporting Information**

*Ravi Teja Addanki Tirumala[⊥], Sundaram Bhardwaj Ramakrishnan[⊥], Farshid Mohammadparast[⊥], Swetha Mahalakshmi Arumugam[§], Susheng Tan[┼], Marimuthu Andiappan\**
*[⊥]*

Affiliations**:**

[⊥] School of Chemical Engineering, Oklahoma State University, Stillwater, OK, USA.

[§] Department of Chemistry, PSG College of Arts & Science, Coimbatore, Tamil Nadu, India.

[┼] Department of Electrical and Computer Engineering and Petersen Institute of Nano Science and Engineering, University of Pittsburgh, Pittsburgh, PA, USA

\* **Corresponding Author**, Email: [mari.andiappan@okstate.edu](mailto:mari.andiappan@okstate.edu)

# I. Syntheses and characterizations of Cu₂O quasi-spherical particles and Cu₂O cubic particles

## Syntheses of Cu₂O nanospheres

Cu₂O quasi-spherical shaped particles, of size 37 ± 6 nm diameter are synthesized using microemulsion technique at room temperature (~20 °C). It is an oil in water emulsion technique where n-heptane (oil phase) is added with polyethylene glycol-dodecyl ether (Brij, average Mn ~362) as surfactant in a 250 mL round bottom flask and allowed to stir at 550 rpm. 0.1 M copper nitrate aqueous solution (5.4 mL) as precursor is added to this mixture and 1 M hydrazine solution (5.4 mL) is added as a reducing agent. Small uniform sizes of reverse micelles formed in mixture acts as a nanoreactor in which the Cu₂O quasi-spherical particles are formed. The mixture is let to stir for 12 hours, after which the emulsion broken using acetone and further centrifuged. These nanoparticles are washed five times (sonicated and centrifuged) to remove the surfactant and separate the nanoparticles for use in DSD experiments.

Cu₂O quasi-spherical particles of 145 ± 41 nm diameter are synthesized using chemical reduction method. 10 mM CuCl₂ aqueous solution (50 mL) as precursor, was prepared in a 100 mL round bottom flask and allowed to stabilize at 55°C, 900 rpm. 2 M NaOH solution (5 mL) was added to the mixture. After 30 minutes, 5 mL of 0.6 M ascorbic acid aqueous solution as reducing agent and the mixture is allowed to stir for 5 hours. The resulting quasi-spherical particles were separated by washing (sonicated and centrifuged) them in DI water and ethanol three times each and used for DSD experiments.

## Syntheses of Cu₂O nanocubes

Cu₂O nanocubes of 165 ± 26 nm edge length were synthesized using a chemical reduction method performed at room temperature (~20 °C). A 250 mL-three-neck round bottom flask (reactor) is flushed with N₂ gas for 30 minutes to make sure there is no oxygen in the reactor atmosphere. A precursor, 120 mL of 0.0032 M aqueous CuCl₂ solution was added to the reactor. 4 mL of 0.35 M aqueous NaOH solution to this mixture. It is observed that mixture turns blue in color due to the formation of Cu(OH)₂ colloids. Add 0.1 M sodium ascorbate of 4 mL, to make 58 mg of small Cu₂O cubes. After which the solution turns light orange in color. The mixture is allowed to stir for 45 minutes, then washed using ethanol three times (sonicate and centrifuge) to obtain 165-nm Cu₂O nanocubes and to use in DSD experiments.

To synthesize Cu₂O nanocubes of 92 ± 13 nm edge length chemical reduction method was used. A 500 mL-three-neck round bottom flask (reactor) at room temperature (~20 °C) is flushed with N₂ gas for 30 minutes to make sure there is no oxygen in the reactor atmosphere. 360 mL of 0.0032 M aqueous CuCl₂ solution as precursor was added to the flask and 12 mL of 0.35 M aqueous NaOH solution is added to this mixture. 0.1 M sodium ascorbate of 12 mL is added to the reactor, after which the solution turns yellowish-orange in color, it is allowed to stir for 45 minutes, after which it is washed using ethanol three times (sonicate and centrifuge) to obtain 92-nm Cu₂O nanocubes.

Chemical reduction method was used for synthesizing $Cu_2O$ cubes of 325 ± 37 nm. 30 mL of 0.0032 M aqueous $CuCl_2$ solution was added to a 50 mL-three neck flask under inert conditions ($N_2$ gas flow). 1 mL of 0.35 M aqueous NaOH solution to this solution at room temperature, after which 0.1 M sodium ascorbate solution is added to the mixture as reducing agent was then added in 1 mL. The solution turns orange in color in the span of 1 hour after which, the $Cu_2O$ nanocubes were washed using ethanol three times (sonicate and centrifuge). 325-nm $Cu_2O$ nanocubes synthesized from this method are further used in DSD experiments.

The synthesized $Cu_2O$ quasi-spherical and cubical particles were characterized using UV-Vis-near IR extinction spectroscopy, X-ray diffraction (XRD) analysis as shown in Figure S1a,b, and transmission electron microscopy (TEM) and scanning electron microscopy (SEM). All UV-Vis-near IR extinction spectra were taken using an Agilent Cary 60 Spectrophotometer. XRD patterns were acquired using a Philips X-Ray diffractometer (Phillips PW 3710 MPD, PW2233/20 X-Ray tube, Copper tube detector – wavelength - 1.5418 Angstroms), operating at 45 KW, 40 mA. The SEM images were taken using an FEI Quanta 600 F. The TEM images were taken using JEOL JEM-2100 TEM and Thermo Fisher Scientific Titan Themis 200 G2 aberration-corrected TEM. The JEOL JEM-2100 system is equipped with a LaB6 gun and an accelerating voltage of 200 kV. The Titan Themis 200 system is equipped with a Schottky field-emission electron gun and operated at 200 kV.

## II. Experimental procedure for performing DSD experiments

In our MB dye-sensitization followed by dye-degradation (DSD) experiments, 6 mL quartz test tube (i.e., photoreactor) is added with 5.8 mg of $Cu_2O$ nanocatalyst (quasi-spherical or cubical nanoparticles) that is uniformly dispersed (sonicated for 2 minutes) in 4 mL of dimethylformamide (DMF). To keep the soluble oxygen ($O_2$) content in the solvent same for all experiments, DMF was sparged with air for 30 minutes. 10 mM methylene blue (MB) solution is made in the DMF from which 150 uL is added to the reaction mixture in the quartz test tube and allowed to stir at 1150 rpm. The photoreactor is shifted to the Luzchem LED Panels (arranged with 4 Luzchem Exposure panels), where Red LED bulbs attached as shown in Figure S1c. Details of intensity measurements are given below. Sampling was done as a function of reaction time by taking 100 µL of the reaction mixture and diluted in 4 mL of DMF. To measure the extent of degradation in DSD experiments with various catalysts $Cu_2O$ spheres and $Cu_2O$ cubes of various sizes, the MB concentration (C) in the reaction mixture was quantified as a function of reaction time. The MB absorption value at its peak absorption wavelength (i.e., 665 nm) was used to quantify the MB concentration in the reaction mixture. The concentration versus stime profiles were fitted to obtain the apparent rate constant values. The fittings were tried with zeroth-order, first-order, and second-order rate equations. Among these trials, the second-order fittings showed the best fit. The fitted apparent rate constant values are provided in Table S1.

The temperature measurements were also done as shown in Figure S3(a-e). Specifically, the temperature of the sample in the photoreactor, temperature of the reactor surrounding, and ambient room temperature of the laboratory are measured. Incident light intensities were measured using Intell Pro Instruments Pro, Smart Sensor purchased from Luzchem Research Inc. The detector is placed exactly where the reactor is placed inside the Luzchem reactor (arranged with 4 Luzchem

Exposure panels). Using the Smart sensor and AR823 Digital Lux meter (i.e., purchased from Luzchem Research Inc), the corresponding settings based on the wavelength range of the LED light intensity are measured in Lux. The values are converted to Light intensity in mW/m$^2$ by multiplying measured lux with the calibration factors. Red light source intensity at the surface of the photoreactor was 6.1375 mW/cm$^2$ and was kept constant for all the experiments reported in this study.

### III. Details of finite-difference time-domain (FDTD) simulations

To compute the extinction spectra of $Cu_2O$, $CeO_2$, $CuO$, $\alpha$-$Fe_2O_3$, and $TiO_2$ of various sizes and shapes, FDTD simulations were used by employing the Lumerical FDTD package.[1] For the calculations and predictions of the dye-sensitization rate enhancements by the $Cu_2O$ nanostructures of different shapes and sizes, we propose a simple approximation, i.e., equation (1) shown in the main draft. This approximation is similar to the approximation proposed by Ingram et al.[2] and used for predicting plasmonic resonance-enhanced photocatalysis. For FDTD simulations, the optical properties (i.e., real and imaginary parts of the refractive index, n and k values) of $CeO_2$, $CuO$, $Cu_2O$, $\alpha$-$Fe_2O_3$, and $TiO_2$ are obtained from the literature.[3-7] The perfectly matched layer (PML) boundary conditions were used for the simulations in all x, y, and z directions. For the simulations of extinction, scattering, and absorption spectra, the respective cross sections as a function of wavelengths were calculated using the total-field/scattered-field (TFSF) formalism. The incident light source used for these simulations was the Gaussian source in the simulated wavelength region. For cubes simulation, the propagation direction of the incident light is perpendicular or parallel to the principal axes. On implementing the simulations of the magnetic and electric field distributions, a plane wave was used as electromagnetic field incidence with propagation in the x-axis direction, and polarization along the y-axis and the z-axis for the electric field and the magnetic field, respectively.

A schematic diagram of the DSD pathway for the dye-only condition is shown in Figure S4a. For the calculations and predictions of the enhancements of the rate of DSD by the $Cu_2O$ nanostructures, we mainly consider here two possible enhancements. The first enhancement (G1) we consider is the enhancement for the light absorption efficiency of the dye molecules. This enhancement can occur via a number of pathways, including the electromagnetic nearfield or nanoantenna effect and Mie resonance-induced resonance energy transfer (MIRET). In our experiments with $Cu_2O$ nanostructures of different shapes and sizes, the weight load of the $Cu_2O$ nanostructures in the reaction mixture was kept constant. For this condition, the enhancement G1 can be written as equation (I) shown below.[2]

$$G1 \propto \int I_0(\lambda) Abs_{MB}(\lambda) E_{Cu2O}(\lambda) \, d\lambda \qquad (I)$$

where $\lambda$ is the wavelength of the incident light, $Ext_{Cu2O}$ is the wavelength-dependent volume-normalized extinction cross section of the $Cu_2O$ nanostructure, $Abs_{MB}$ is the wavelength-dependent absorbance of MB, and $I_0$ is the wavelength-dependent intensity of the incident light.

The second enhancement (G2) we consider is the enhancement by the catalytic surface effect of $Cu_2O$ nanostructures that can also facilitate and enhance the electron transfer between adsorbed MB and $O_2$ molecules and the subsequent degradation of the excited MB molecules. This enhancement (G2) will be proportional to the ratio of the surface area to the volume of the $Cu_2O$ nanostructures (S/V). The rate of DSD in the presence of $Cu_2O$ nanostructures will be proportional to the overall enhancement, which is the product of these two enhancements (G1 and G2). Based on this overall enhancement, the rate (r) of DSD in the presence of $Cu_2O$ nanostructures can be written as equation (1) shown below.

$$r \propto \int I_0(\lambda) Abs_{MB}(\lambda) E_{Cu2O}(\lambda) d\lambda \times \frac{S}{V} \qquad (1)$$

In the above equation, the volume-normalized extinction cross section ($E_{Cu2O}$) is the ratio of extinction cross section of the $Cu_2O$ nanostructure to its volume ($\sigma_{Ext}/V$). Therefore, $\sigma_{Ext}/V$ will be proportional to $A^{-3}$ for nanocubes (where A is the edge length), and $D^{-3}$ for the nanospheres (where D is the diameter). The surface to volume (S/V) ratio in the above equation will be proportional to the $A^{-1}$ for nanocubes, and $D^{-1}$ for the nanospheres. Therefore, the extinction cross section value normalized to the fourth power of the edge length ($\sigma_{Ext}/A^4$, for nanocubes) or the fourth power of diameter (($\sigma_{Ext}/D^4$, for nanospheres) would be a good descriptor to predict the relative rate of DSD on $Cu_2O$ nanostructures of different sizes. For example, the normalized extinction value ($\sigma_{Ext}/A^4$) in Figure 4a is a good descriptor to predict the relative rate of DSD on $Cu_2O$ nanocubes of different sizes.

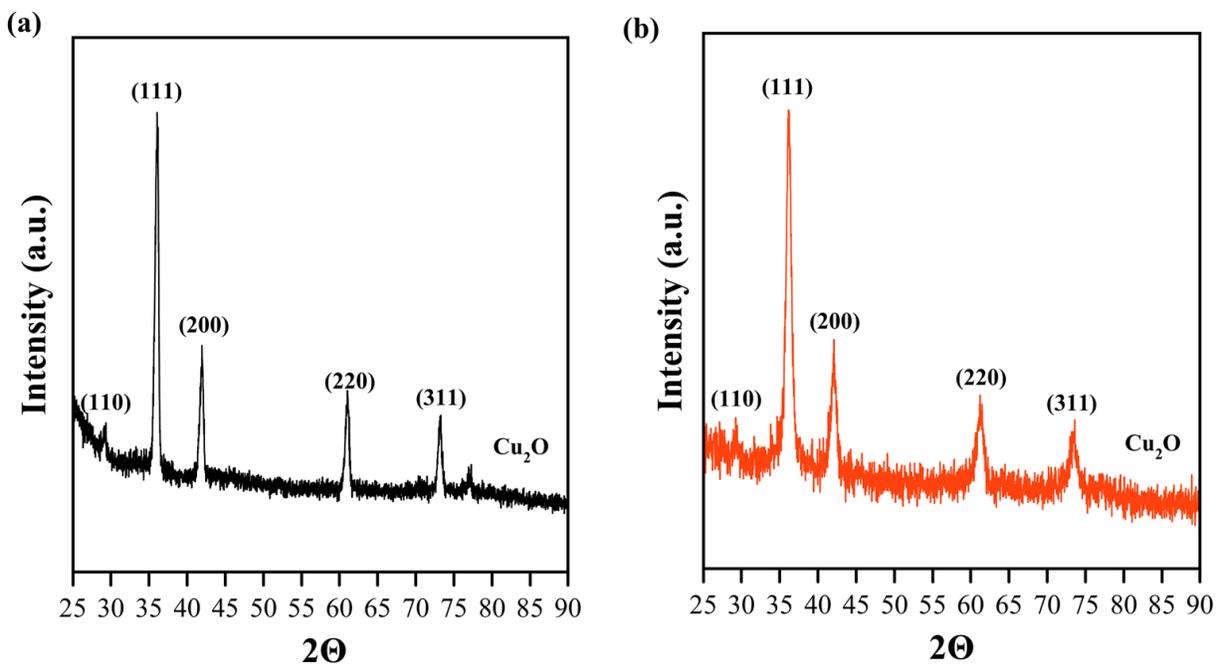

**Figure S1.** Representative XRD spectra of **(a)** Cu₂O nanospheres and **(b)** Cu₂O nanocubes

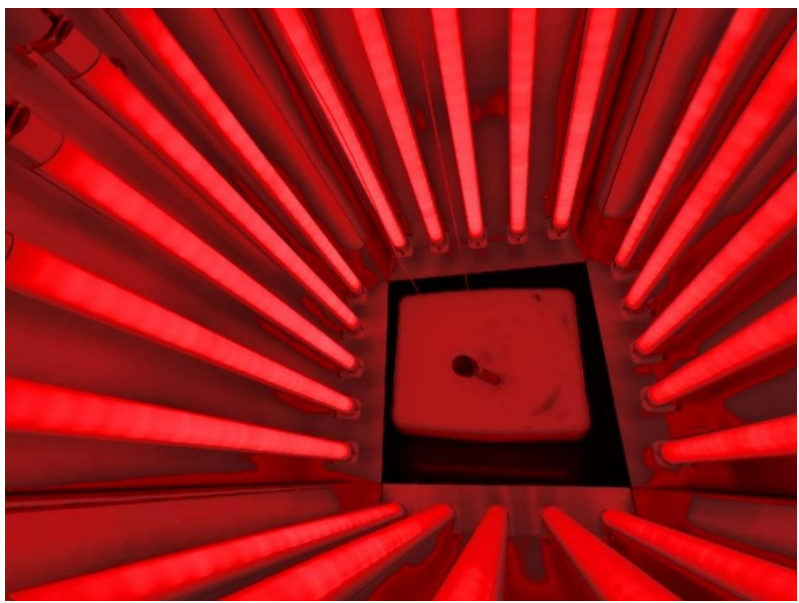

**Figure S1 (c)** Luzchem photoreactor experimental set up with red light illumination used for dye sensitization studies.

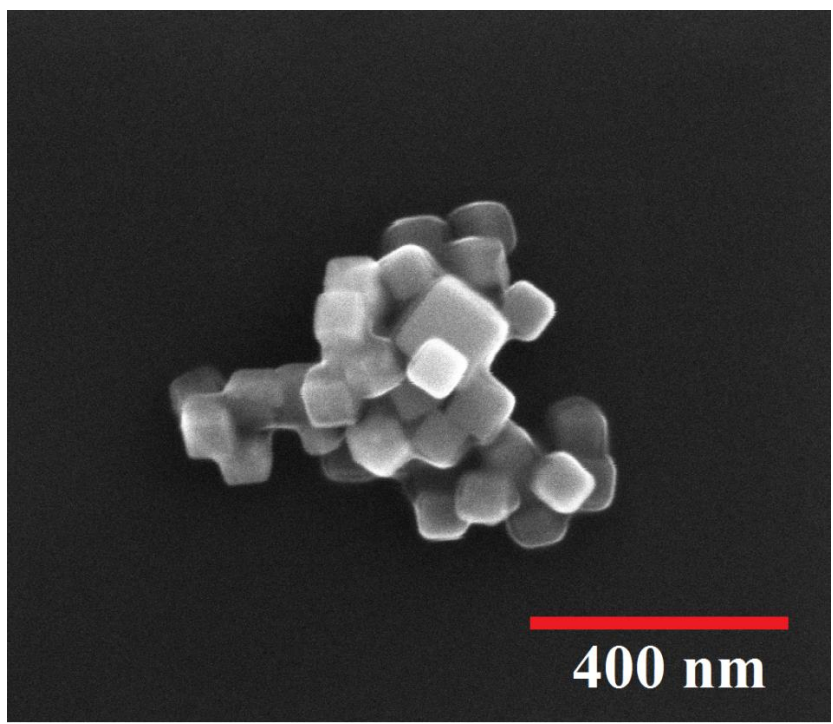

**Figure S1 (d)** Representative transmission electron microcopy image of large Cu₂O nanocubes of 92 ± 13 nm edge length synthesized using chemical reduction method.

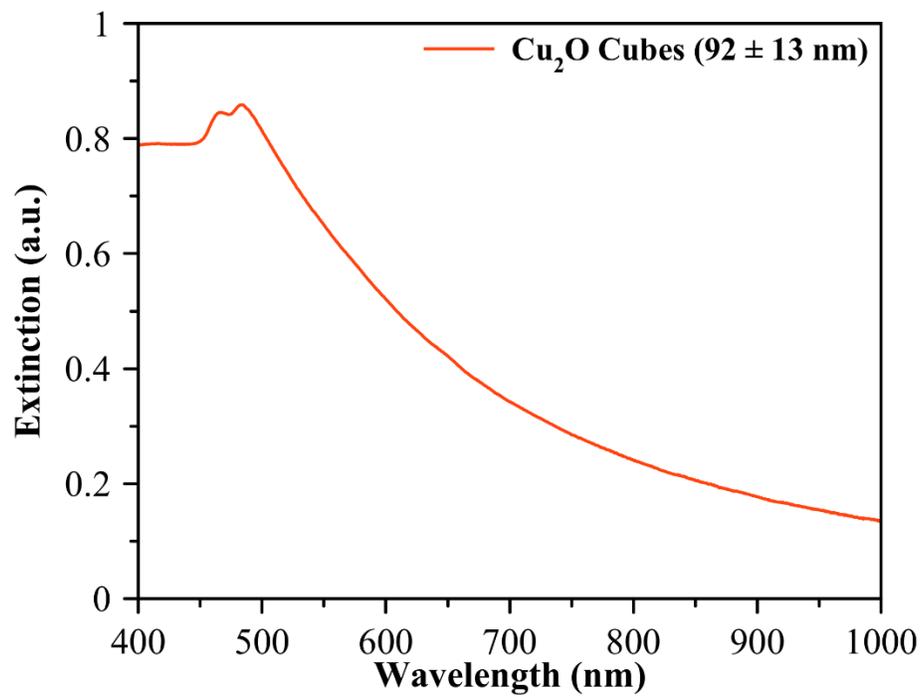

**Figure S1 (e)** Experimentally measured UV-Vis extinction spectra of Cu₂O nanocubes of 92 ± 13 nm edge length.

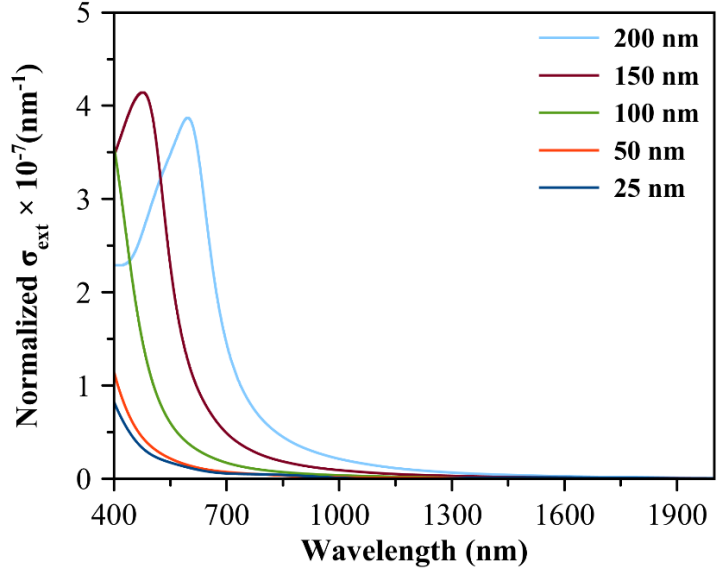

**Figure S2 (a)** FDTD-simulated normalized-extinction (normalized-$\sigma_{Ext}$) cross section of $Cu_2O$ nanospheres of different edge lengths in the range of 25 to 200 nm as a function of incident light wavelength. Normalized-$\sigma_{Ext}$ = Ratio of extinction cross section ($\sigma_{Ext}$) to the volume of the nanosphere.

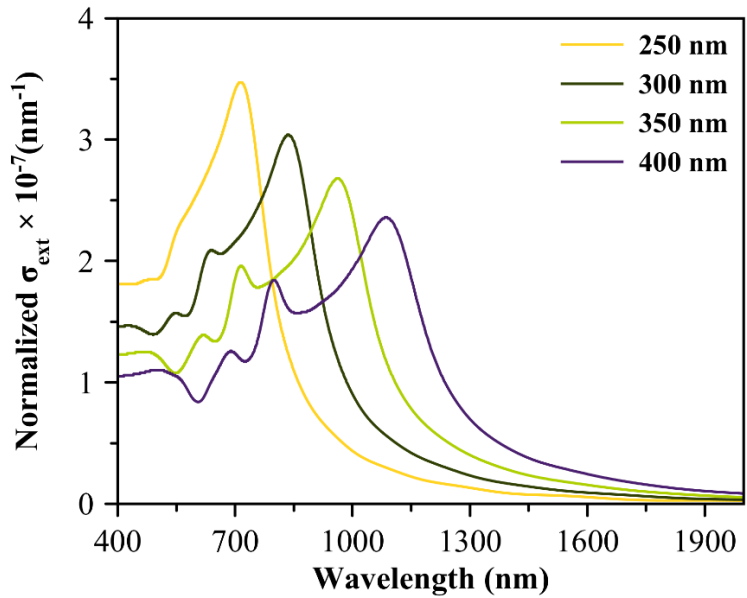

**Figure S2 (b)** FDTD-simulated normalized-extinction (normalized-$\sigma_{Ext}$) cross section of $Cu_2O$ nanospheres of different edge lengths in the range of 250 to 400 nm as a function of incident light wavelength. Normalized-$\sigma_{Ext}$ = Ratio of extinction cross section ($\sigma_{Ext}$) to the volume of the nanosphere.

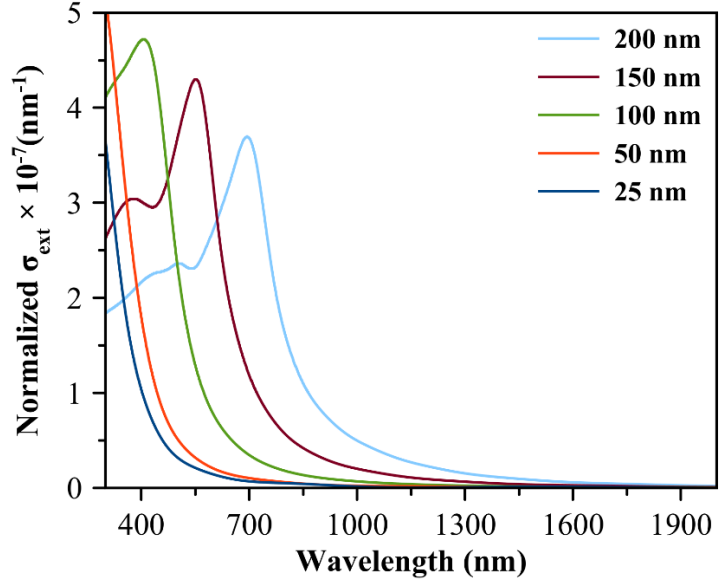

**Figure S2 (c)** FDTD-simulated normalized-extinction (normalized-$\sigma_{Ext}$) cross section of $Cu_2O$ nanocubes of different edge lengths in the range of 25 to 200 nm as a function of incident light wavelength. Normalized-$\sigma_{Ext}$ = Ratio of extinction cross section ($\sigma_{Ext}$) to the volume of the nanocube.

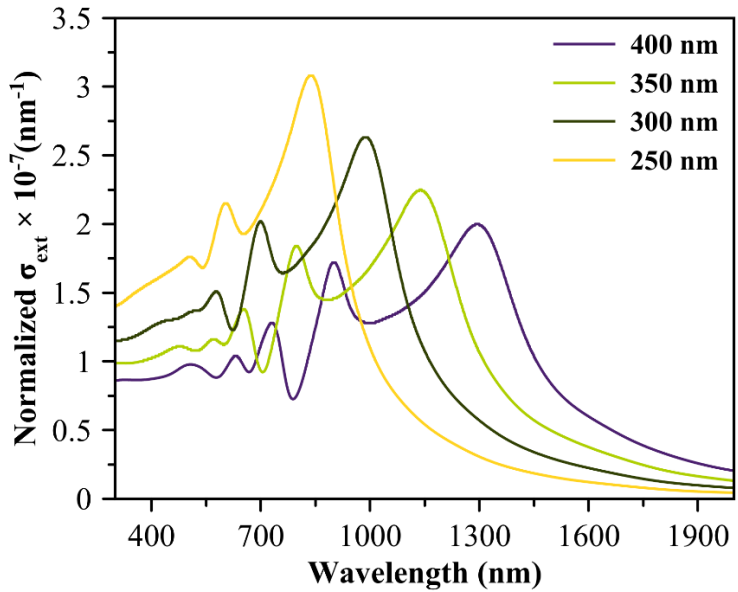

**Figure S2 (d)** FDTD-simulated normalized-extinction (normalized-$\sigma_{Ext}$) cross section of $Cu_2O$ nanocubes of different edge lengths in the range of 250 to 400 nm as a function of incident light wavelength. Normalized-$\sigma_{Ext}$ = Ratio of extinction cross section ($\sigma_{Ext}$) to the volume of the nanocube.

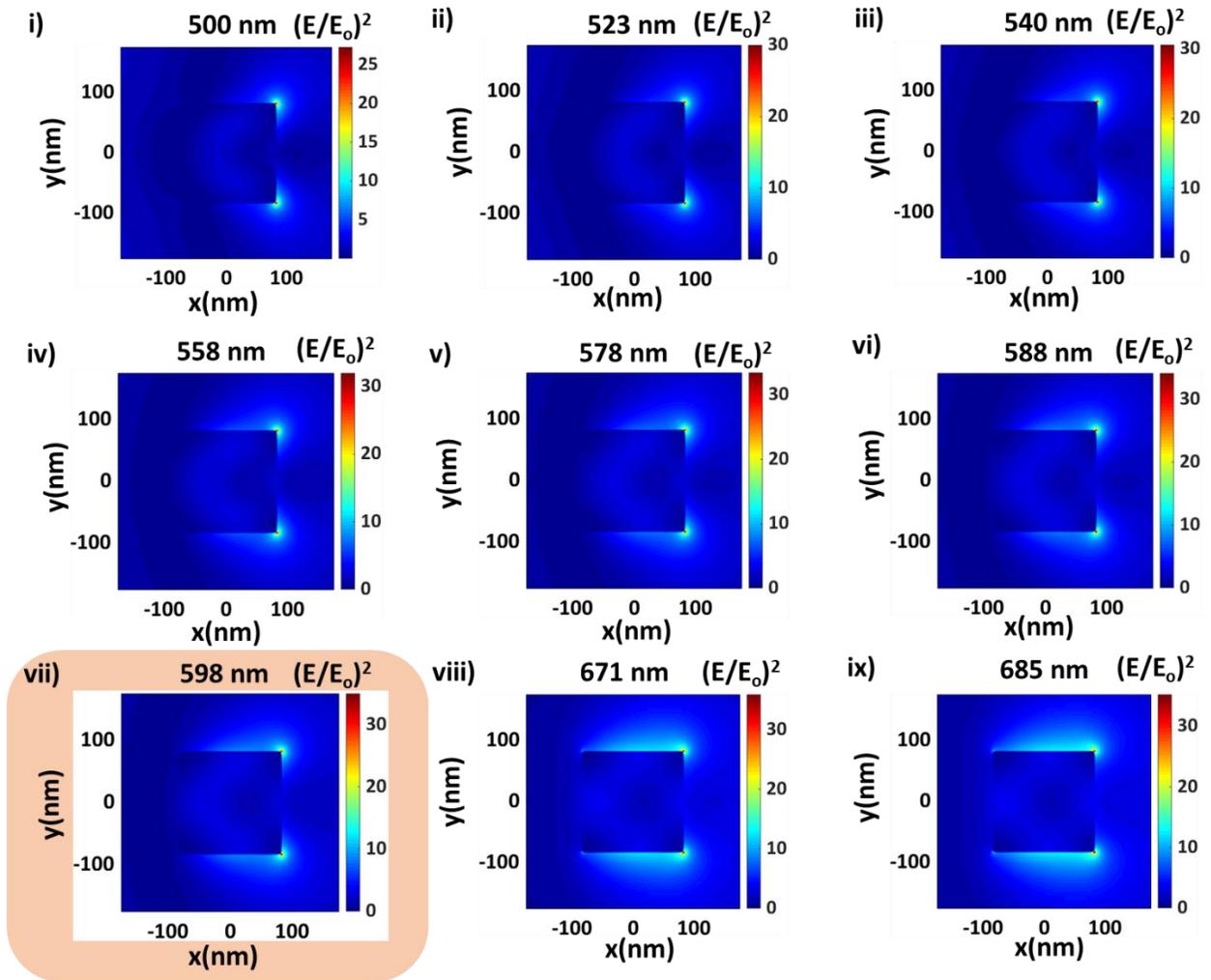

**Figure S2 (e) i.** Simulated spatial distribution of enhancement in electric field intensity [$E^2/E_0^2$] in XY plane at different wavelengths across the Mie resonance peak wavelength (i.e., **598 nm**) for Cu$_2$O nanocubes with an edge length of 165 nm.

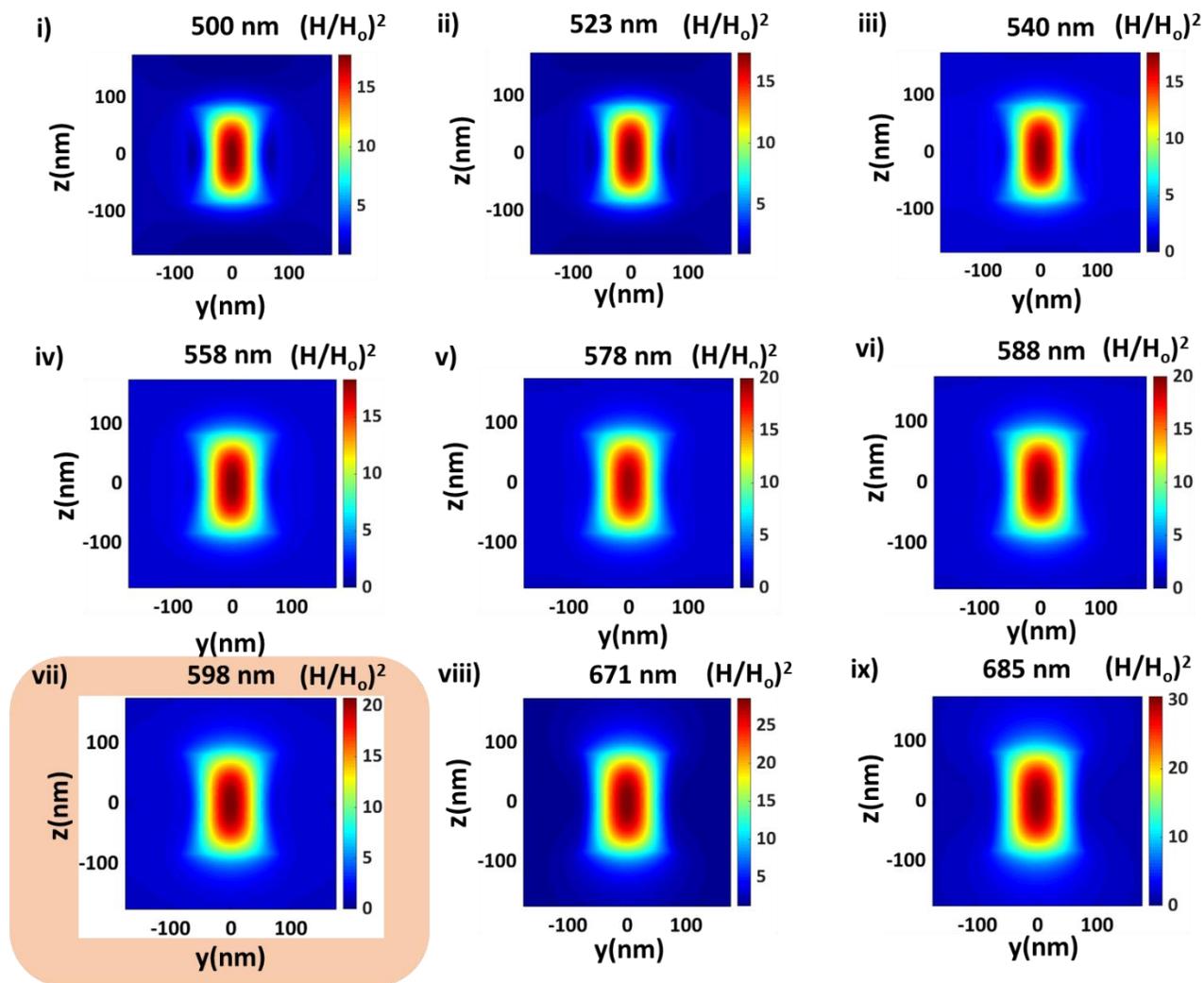

**Figure S2 (e) ii.** Simulated spatial distribution of enhancement in magnetic field intensity [$H^2/H_0^2$] in YZ plane at different wavelengths across the Mie resonance peak wavelength (i.e., **598 nm**) for $Cu_2O$ nanocubes with an edge length of 165 nm.

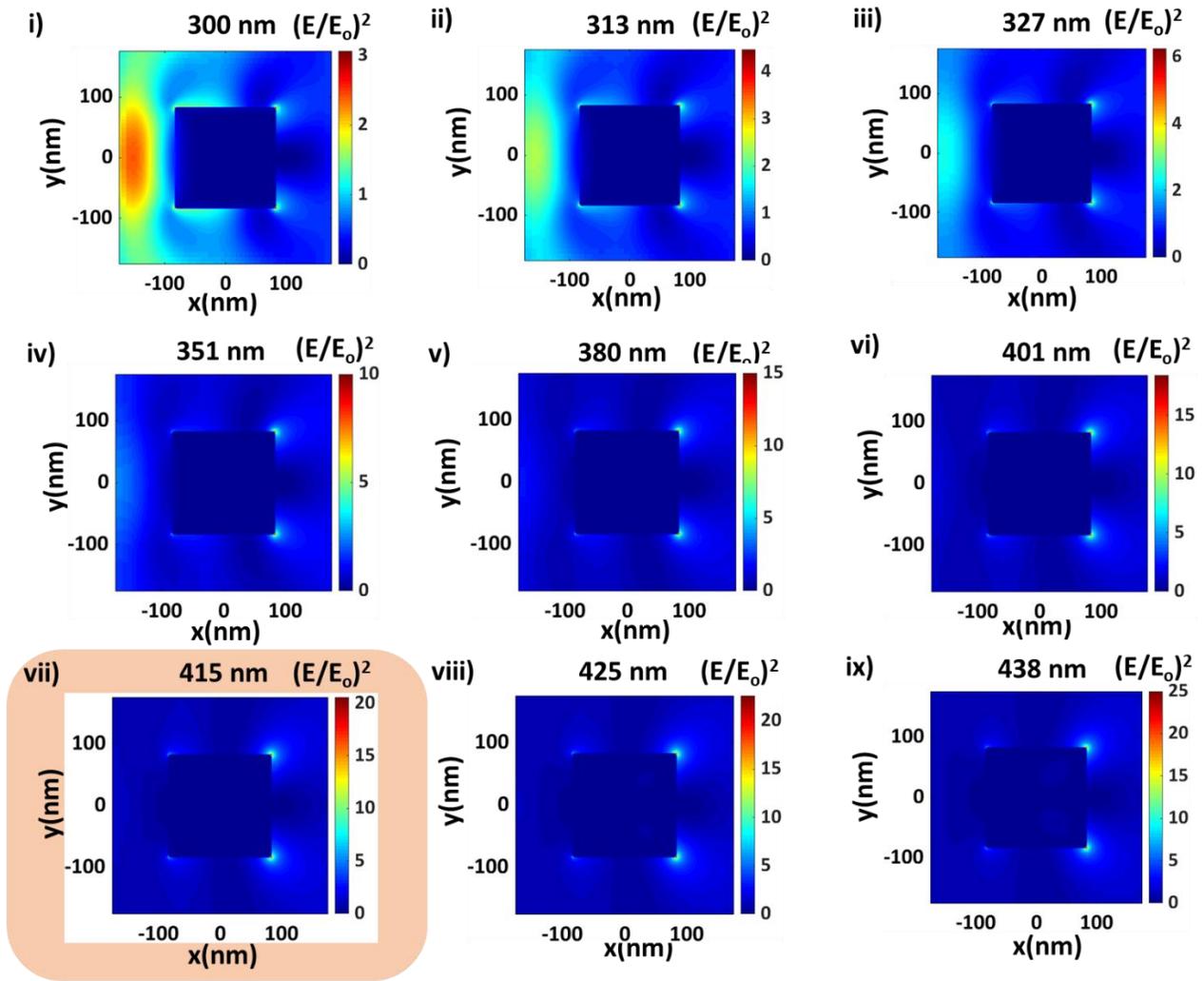

**Figure S2 (f) i.** Simulated spatial distribution of enhancement in electric field intensity [$E^2/E_0^2$] in XY plane at different wavelengths across the Mie resonance peak wavelength (i.e., **415 nm**) for Cu$_2$O nanocubes with an edge length of 165 nm.

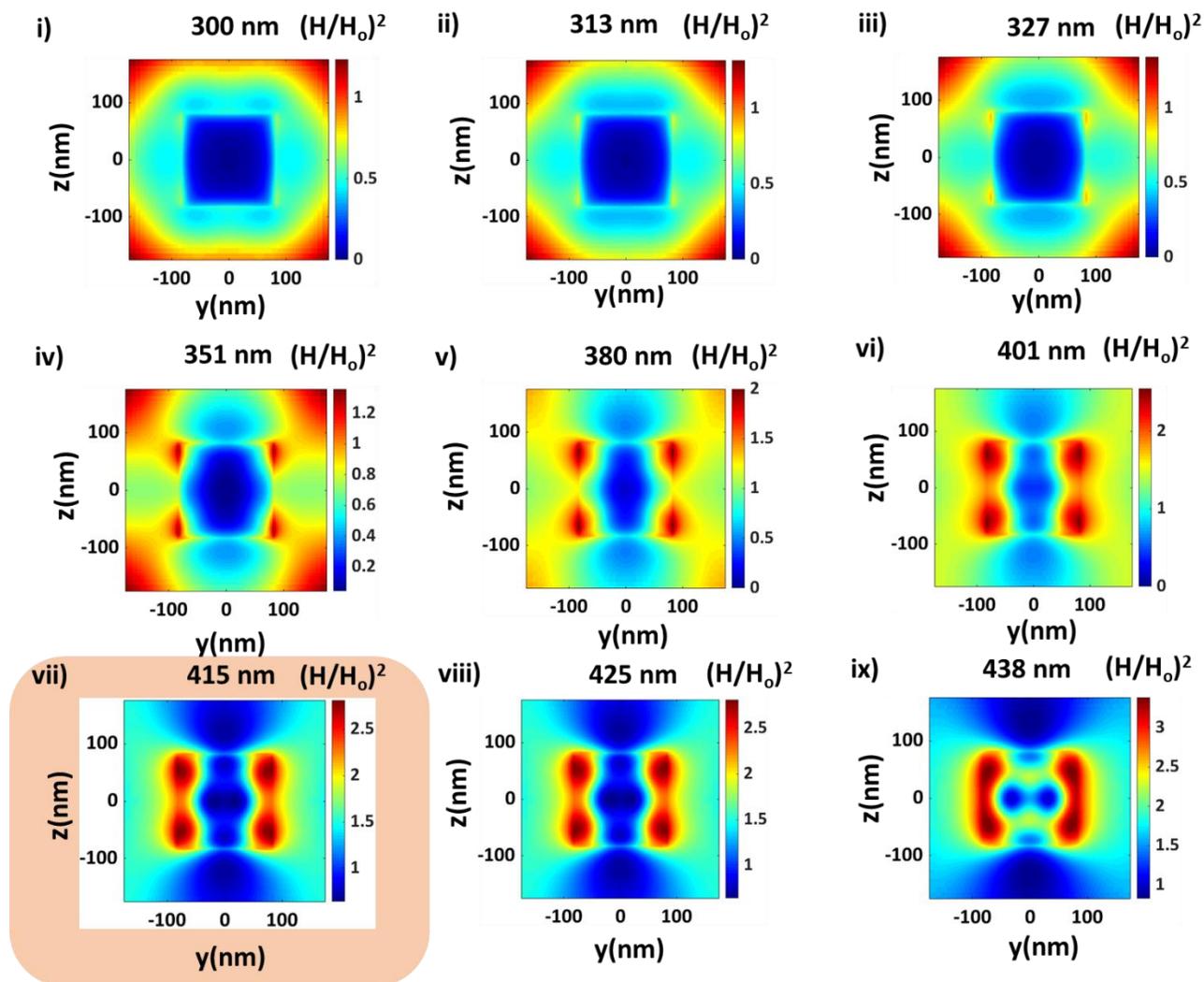

**Figure S2 (f) ii.** Simulated spatial distribution of enhancement in magnetic field intensity [$H^2/H_0^2$] in YZ plane at different wavelengths across the Mie resonance peak wavelength (i.e., **415 nm**) for $Cu_2O$ nanocubes with an edge length of 165 nm.

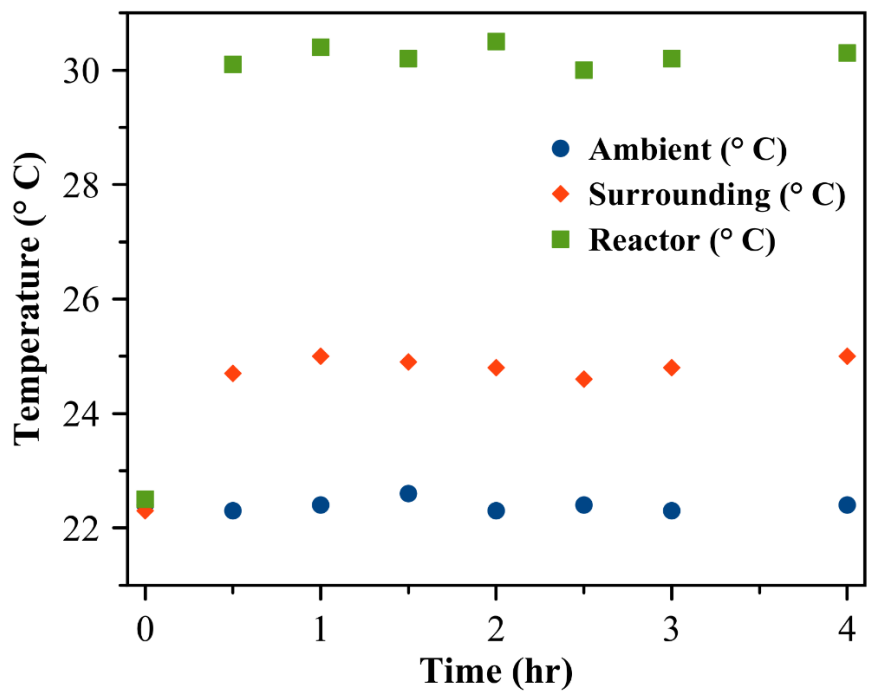

**Figure S3(a).** Temperature profile measured as a function of irradiation time for MB DSD in DMF using quasi-spherical $Cu_2O$ nanoparticles of $37 \pm 6$ nm diameter.

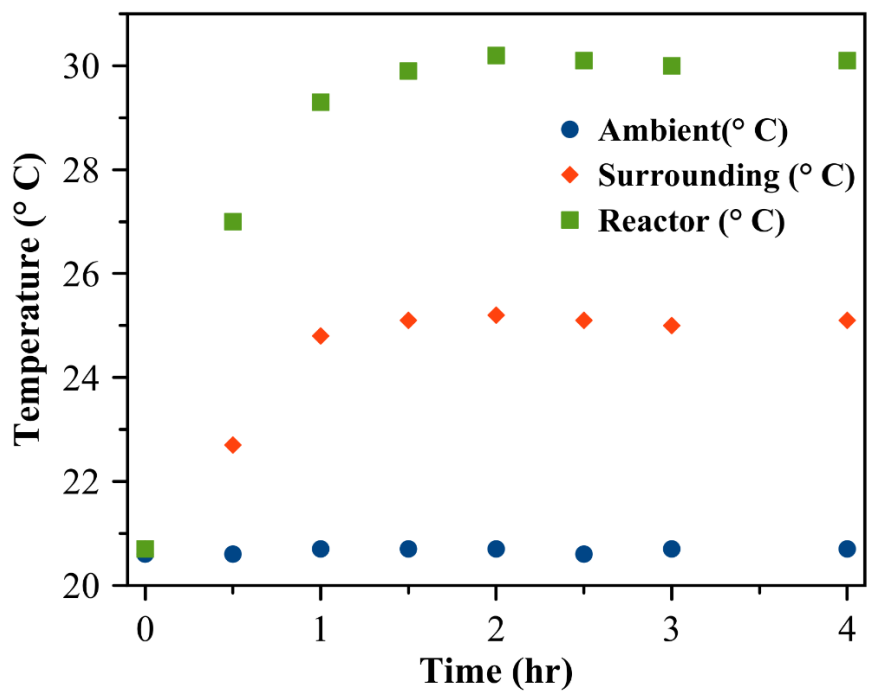

**Figure S3(b).** Temperature profile measured as a function of irradiation time for MB DSD in DMF using quasi-spherical $Cu_2O$ nanoparticles of $145 \pm 41$ nm diameter.

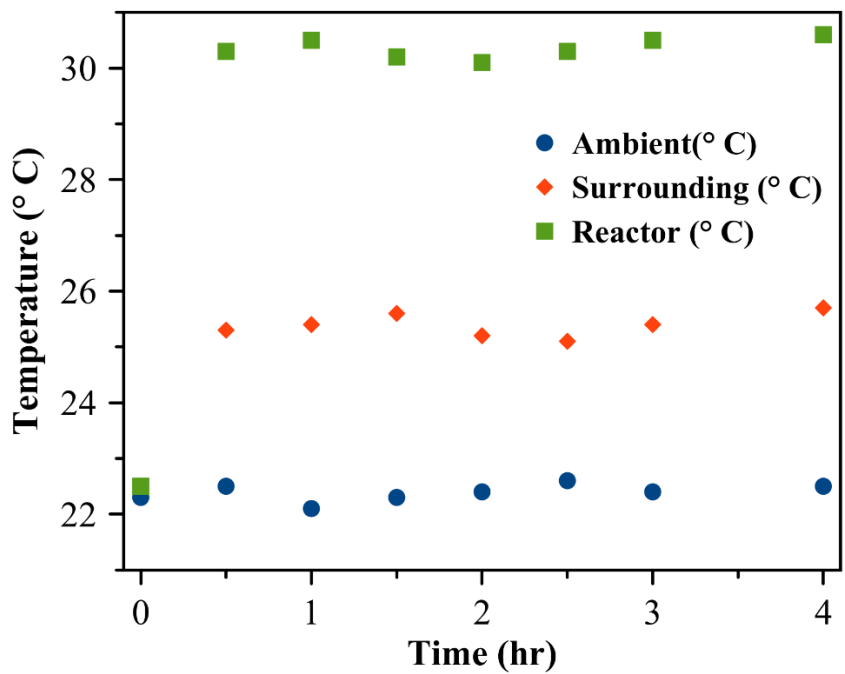

**Figure S3(c).** Temperature profile measured as a function of irradiation time for MB DSD in DMF using Cu$_2$O nanocubes of 165 ± 26 nm edge length.

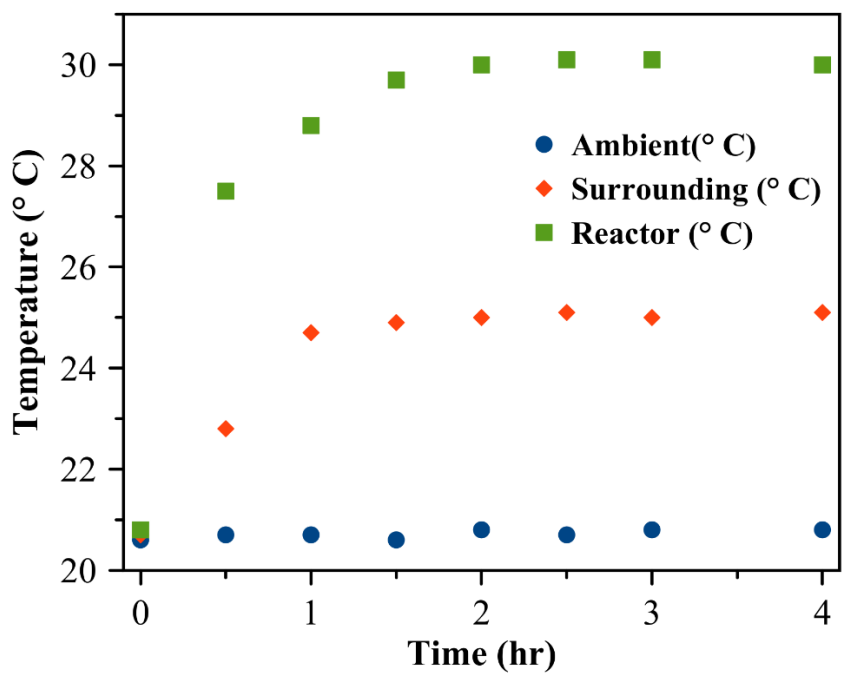

**Figure S3(d).** Temperature profile measured as a function of irradiation time for MB DSD in DMF using Cu$_2$O nanocubes of 325 ± 37 nm edge length.

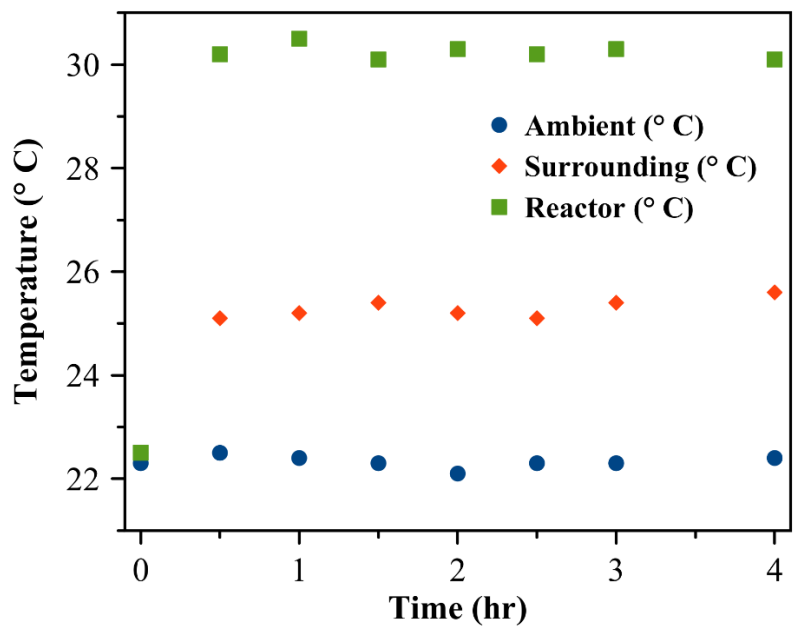

**Figure S3(e).** Temperature profile measured as a function of irradiation time for MB DSD in DMF under blank conditions in the absence of photocatalyst.

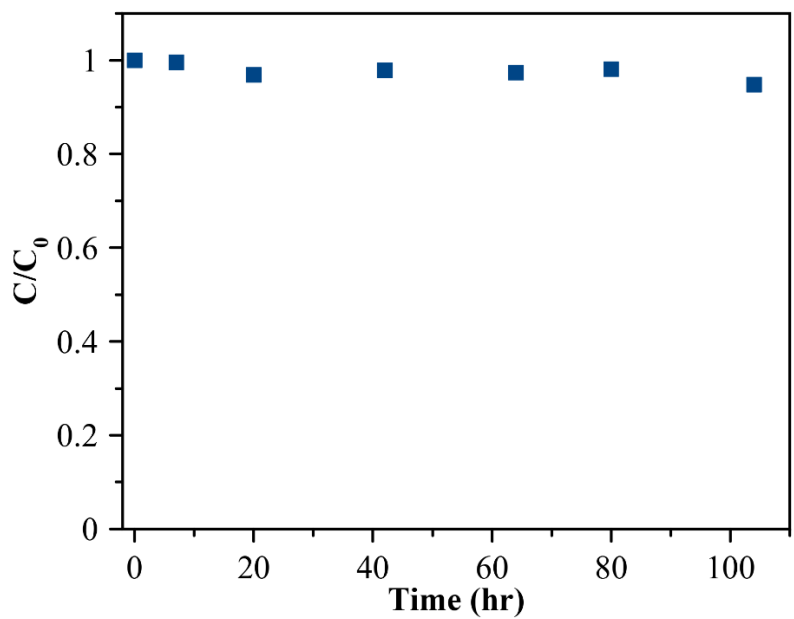

**Figure S3(f).** Plot of $C/C_0$ of methylene blue versus reaction time measured during heating experiments at 60 °C in the presence of large $Cu_2O$ cubes for MB DSD under dark conditions in the absence of red-light irradiation.

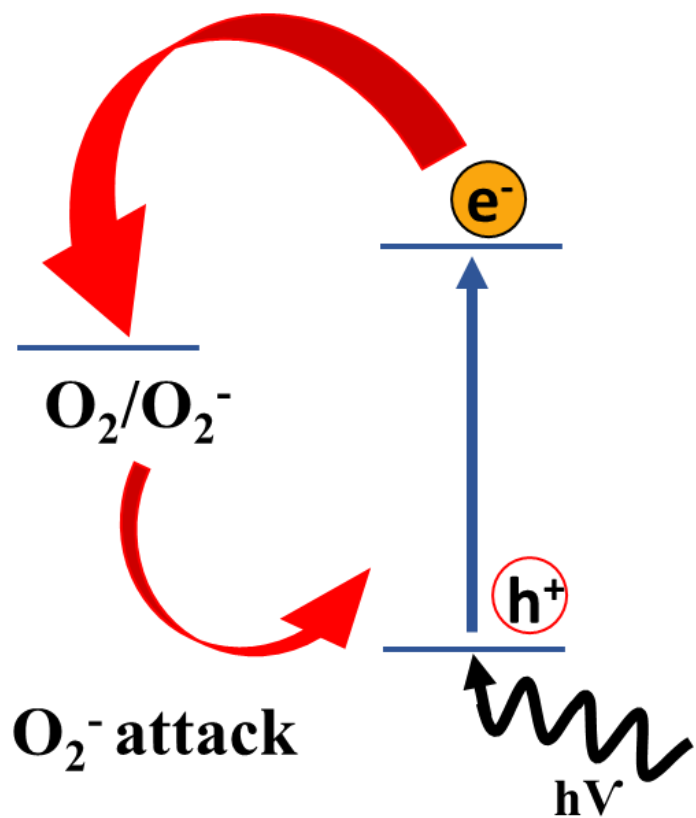

**Figure S4 (a)** Schematic diagram of dye-only dye degradation pathway. Dissolved oxygen in the solvent can capture the excited electron from the dye to form superoxide ($O_2^-$) radical, which can attack the dye and degrade it. (Reference: Rochkind, M.; Pasternak, S.; Paz, Y. Using Dyes for Evaluating Photocatalytic Properties: A Critical Review. Molecules 2015, 20 (1), 88–110)

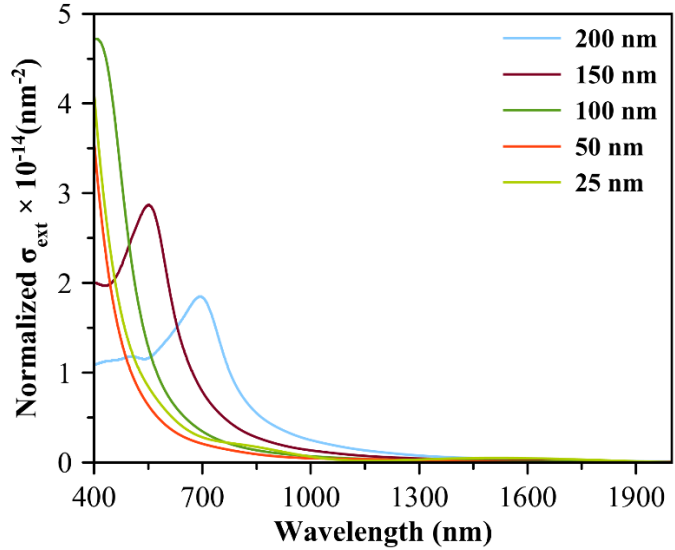

**Figure S4 (b)** FDTD-simulated normalized-extinction (normalized-$\sigma_{Ext}$) cross section of $Cu_2O$ nanocubes of different edge lengths in the range of 25 to 200 nm as a function of incident light wavelength.

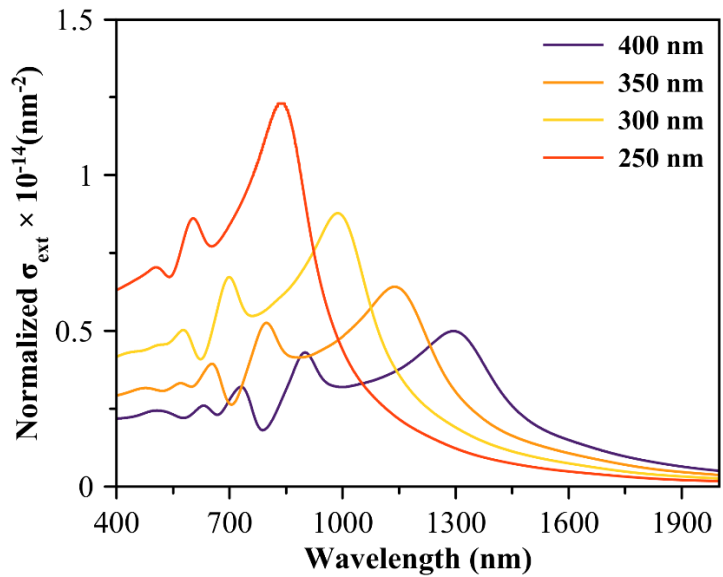

**Figure S4 (c)** FDTD-simulated normalized-extinction (normalized-$\sigma_{Ext}$) cross section of $Cu_2O$ nanocubes of different edge lengths in the range of 250 to 400 nm as a function of incident light wavelength.

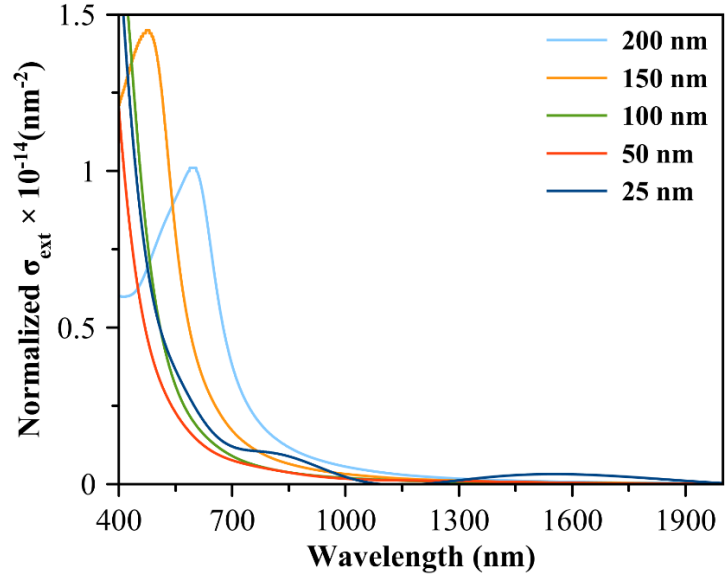

**Figure S4 (d)** FDTD-simulated normalized-extinction (normalized-$\sigma_{Ext}$) cross section of $Cu_2O$ nanospheres of different edge lengths in the range of 25 to 200 nm as a function of incident light wavelength

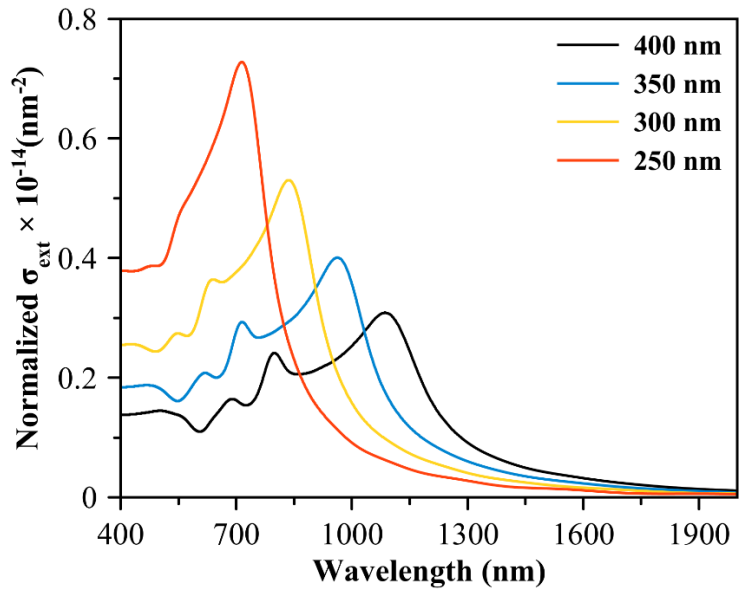

**Figure S4 (e)** FDTD-simulated normalized-extinction (normalized-$\sigma_{Ext}$) cross section of $Cu_2O$ nanospheres of different edge lengths in the range of 250 to 400 nm as a function of incident light wavelength.

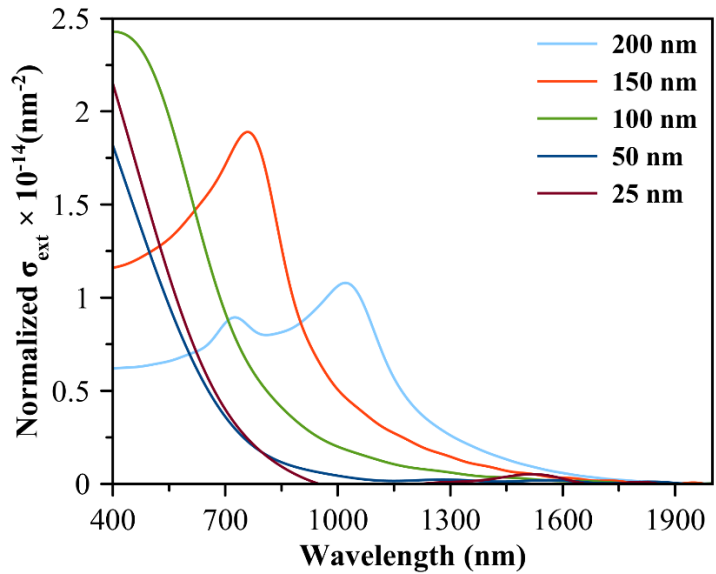

**Figure S4 (f)** FDTD-simulated normalized-extinction (normalized-$\sigma_{Ext}$) cross section of $CeO_2$ nanospheres of different diameter in the range of 25 to 200 nm as a function of incident light wavelength.

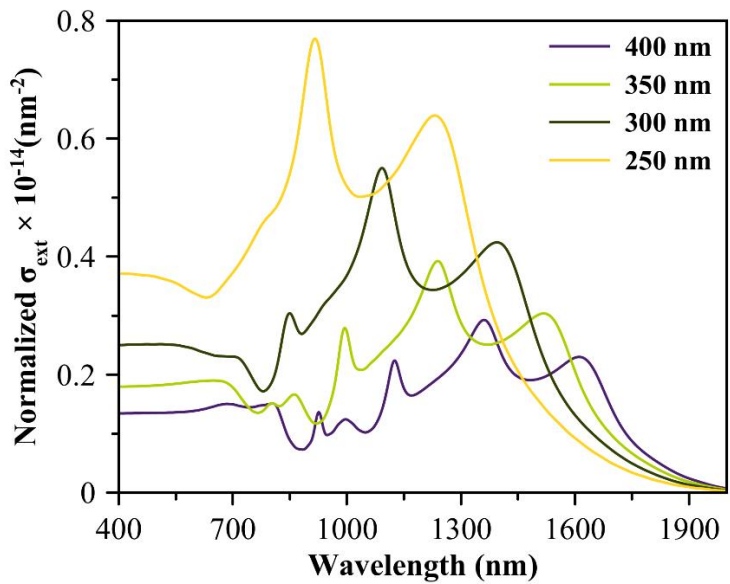

**Figure S4 (g)** FDTD-simulated normalized-extinction (normalized-$\sigma_{Ext}$) cross section of $CeO_2$ nanospheres of different diameter in the range of 250 to 400 nm as a function of incident light wavelength.

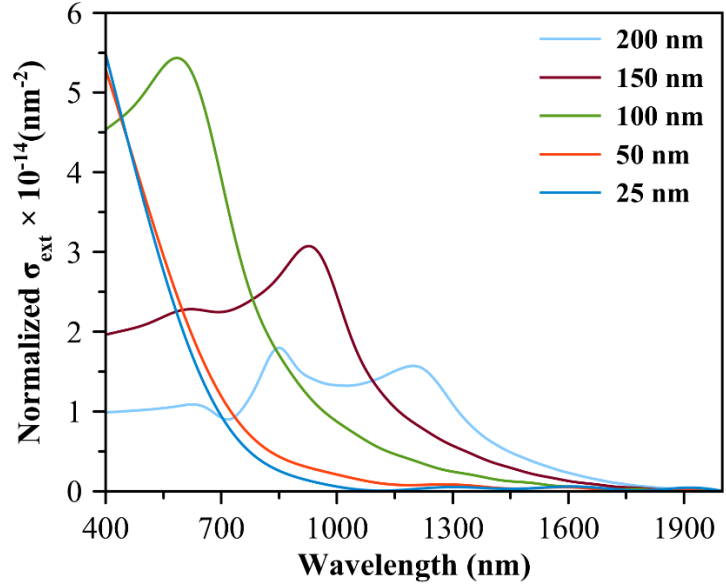

**Figure S4 (h)** FDTD-simulated normalized-extinction (normalized-$\sigma_{Ext}$) cross section of CeO$_2$ nanocubes of different edge length in the range of 25 to 200 nm as a function of incident light wavelength.

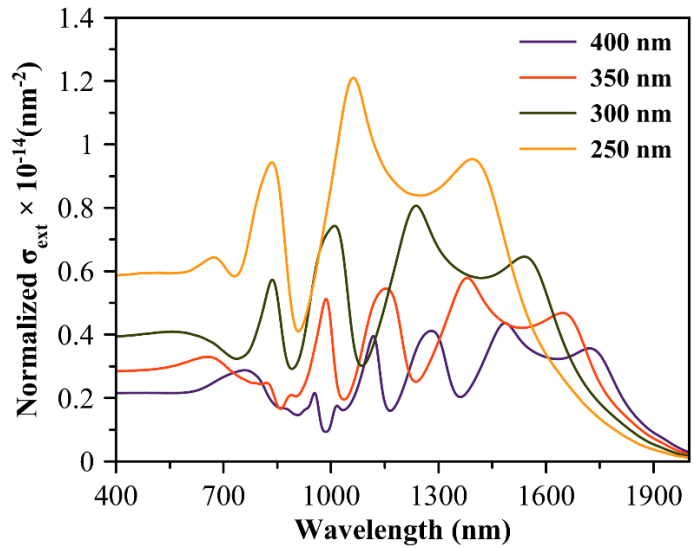

**Figure S4 (i)** FDTD-simulated normalized-extinction (normalized-$\sigma_{Ext}$) cross section of CeO$_2$ nanocubes of different edge length in the range of 250 to 400 nm as a function of incident light wavelength.

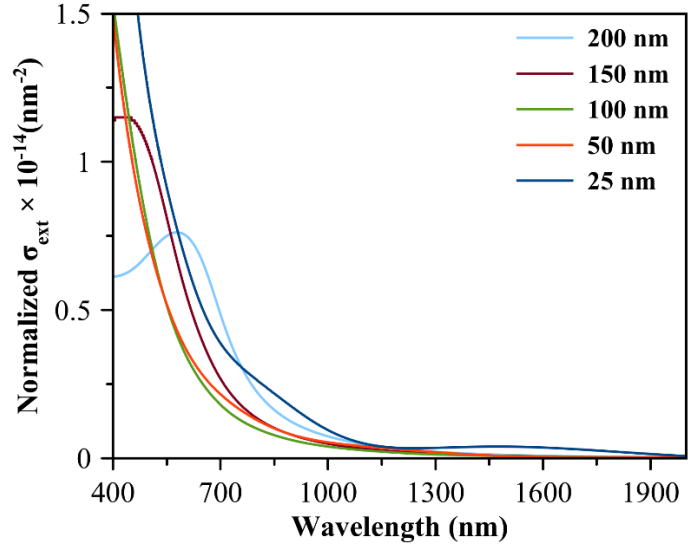

**Figure S4 (j)** FDTD-simulated normalized-extinction (normalized-$\sigma_{Ext}$) cross section of CuO nanospheres of different diameter in the range of 25 to 200 nm as a function of incident light wavelength.

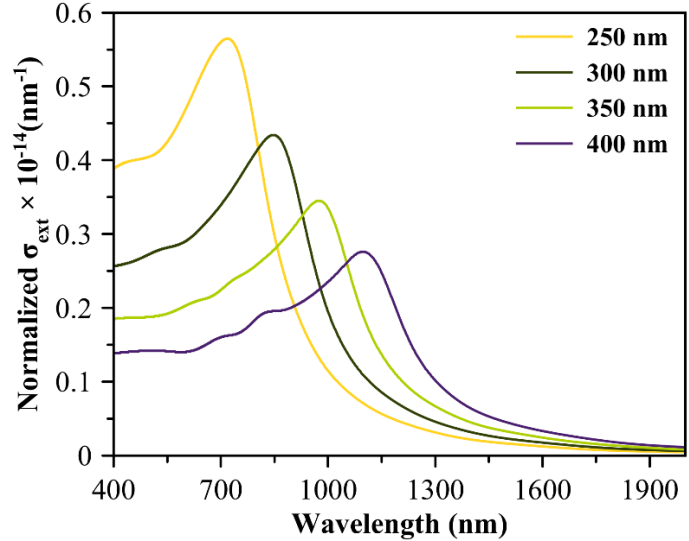

**Figure S4 (k)** FDTD-simulated normalized-extinction (normalized-$\sigma_{Ext}$) cross section of CuO nanospheres of different diameter in the range of 250 to 400 nm as a function of incident light wavelength.

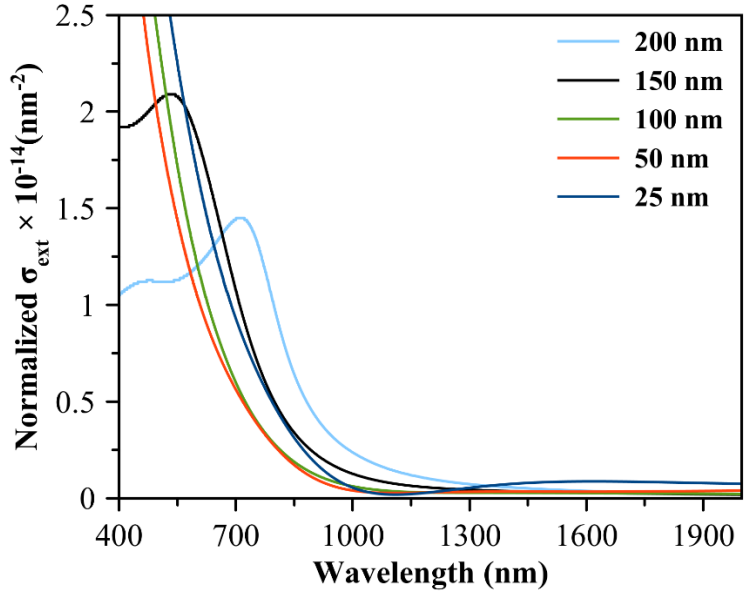

**Figure S4 (l)** FDTD-simulated normalized-extinction (normalized-$\sigma_{Ext}$) cross section of CuO nanocubes of different edge length in the range of 25 to 200 nm as a function of incident light wavelength.

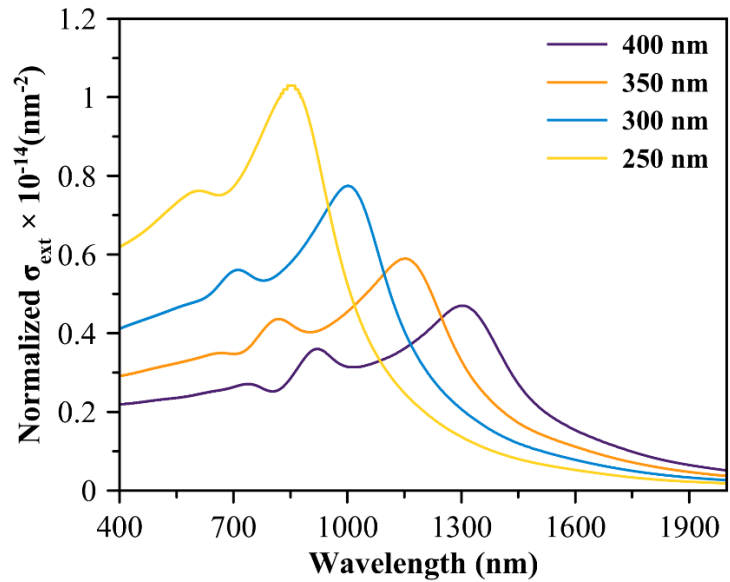

**Figure S4 (m)** FDTD-simulated normalized-extinction (normalized-$\sigma_{Ext}$) cross section of CuO nanocubes of different edge length in the range of 250 to 400 nm as a function of incident light wavelength.

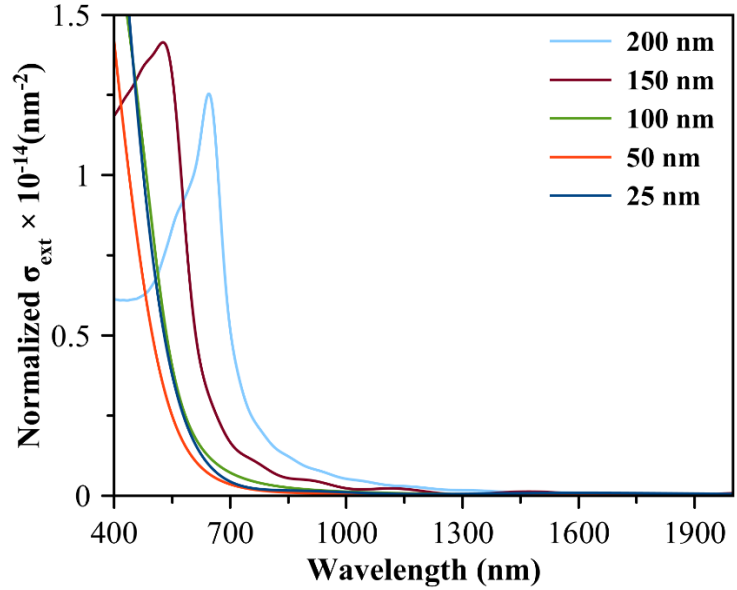

**Figure S4 (n)** FDTD-simulated normalized-extinction (normalized-$\sigma_{Ext}$) cross section of α-Fe$_2$O$_3$ nanospheres of different diameter in the range of 25 to 200 nm as a function of incident light wavelength.

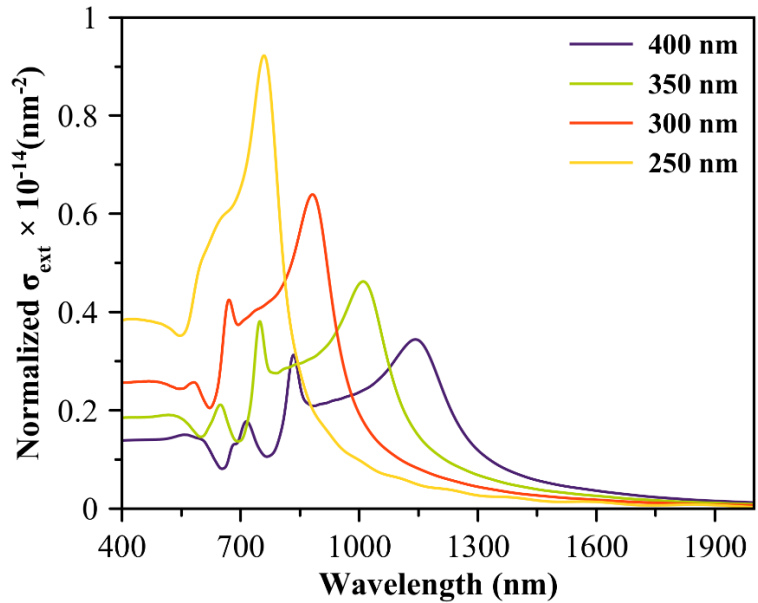

**Figure S4 (o)** FDTD-simulated normalized-extinction (normalized-$\sigma_{Ext}$) cross section of α-Fe$_2$O$_3$ nanospheres of different diameter in the range of 250 to 400 nm as a function of incident light wavelength.

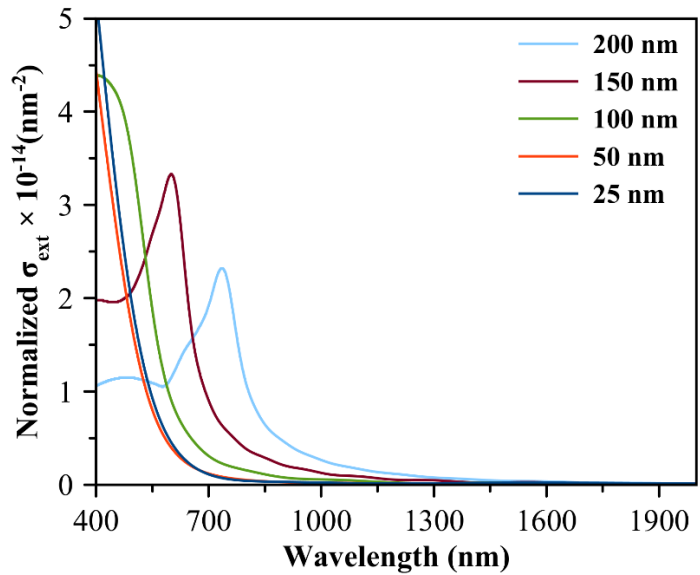

**Figure S4 (p)** FDTD-simulated normalized-extinction (normalized-$\sigma_{Ext}$) cross section of α-Fe$_2$O$_3$ nanocubes of different edge length in the range of 25 to 200 nm as a function of incident light wavelength.

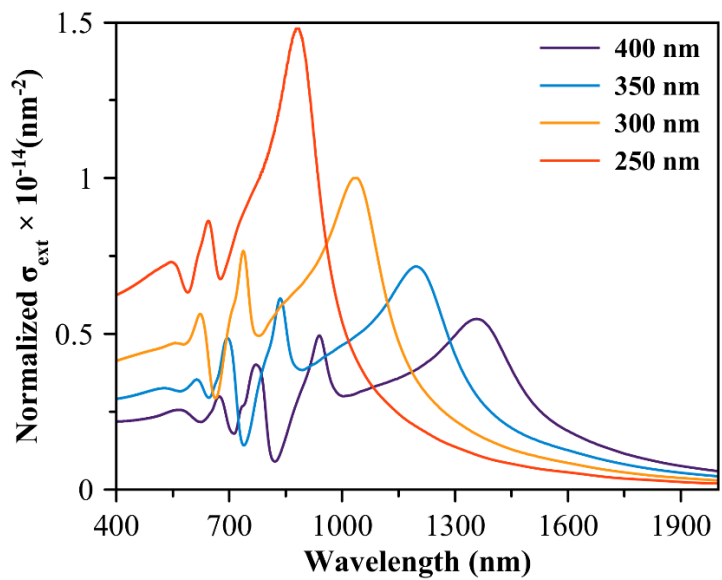

**Figure S4 (q)** FDTD-simulated normalized-extinction (normalized-$\sigma_{Ext}$) cross section of α-Fe$_2$O$_3$ nanocubes of different edge length in the range of 250 to 400 nm as a function of incident light wavelength.

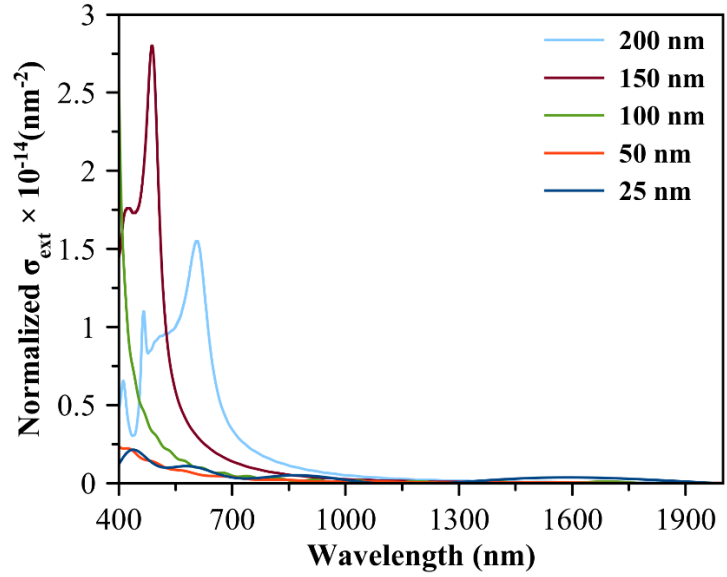

**Figure S4 (r)** FDTD-simulated normalized-extinction (normalized-$\sigma_{Ext}$) cross section of TiO$_2$ nanospheres of different diameter in the range of 25 to 200 nm as a function of incident light wavelength.

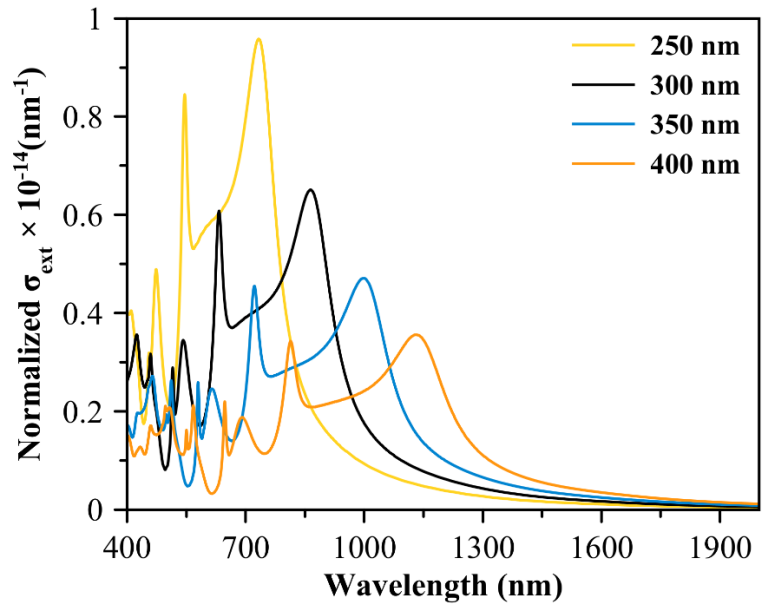

**Figure S4 (s)** FDTD-simulated normalized-extinction (normalized-$\sigma_{Ext}$) cross section of TiO$_2$ nanospheres of different diameter in the range of 250 to 400 nm as a function of incident light wavelength.

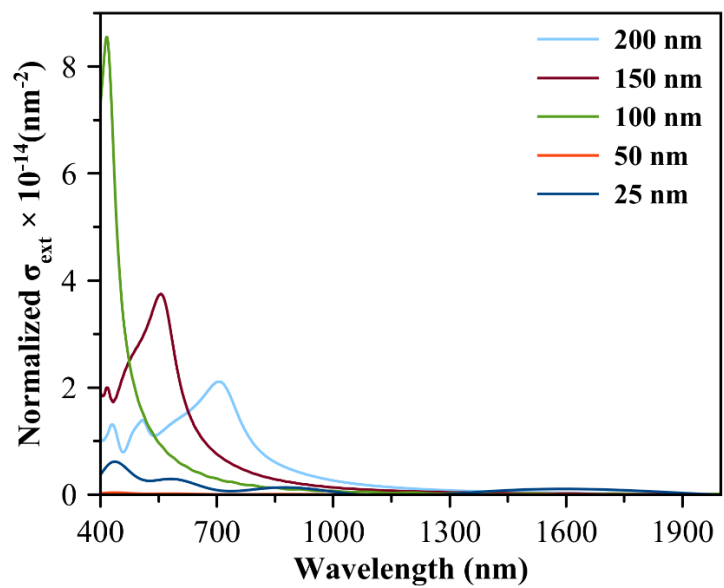

**Figure S4 (t)** FDTD-simulated normalized-extinction (normalized-$\sigma_{Ext}$) cross section of $TiO_2$ nanocubes of different edge length in the range of 25 to 200 nm as a function of incident light wavelength.

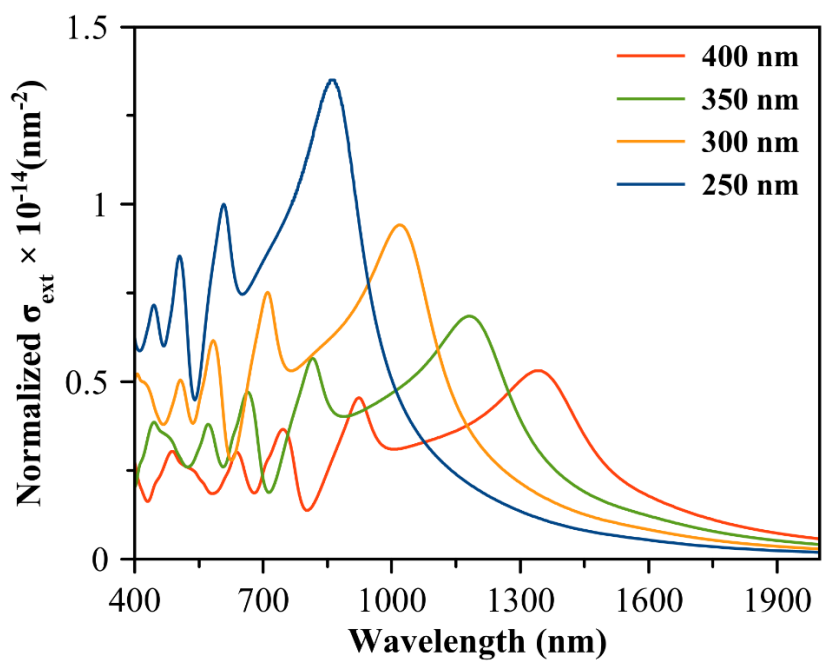

**Figure S4 (u)** FDTD-simulated normalized-extinction (normalized-$\sigma_{Ext}$) cross section of $TiO_2$ nanocubes of different edge length in the range of 250 to 400 nm as a function of incident light wavelength.

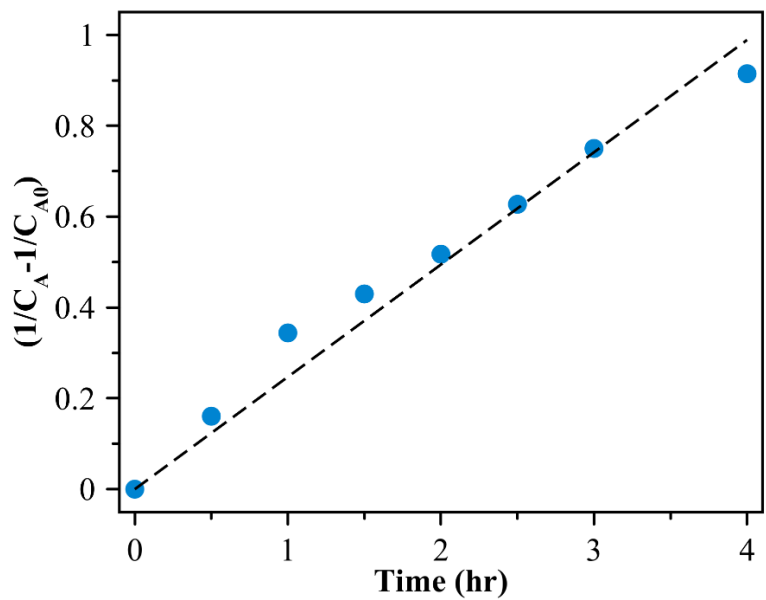

**Figure S4 (v)** Plot of Empirical fit with second order $(1/C_A-1/C_{A0})$ versus irradiation time for MB DSD in DMF, using quasi-spherical $Cu_2O$ nanoparticles of $37 \pm 6$ nm diameter and corresponding trendline and proportion of the variance.

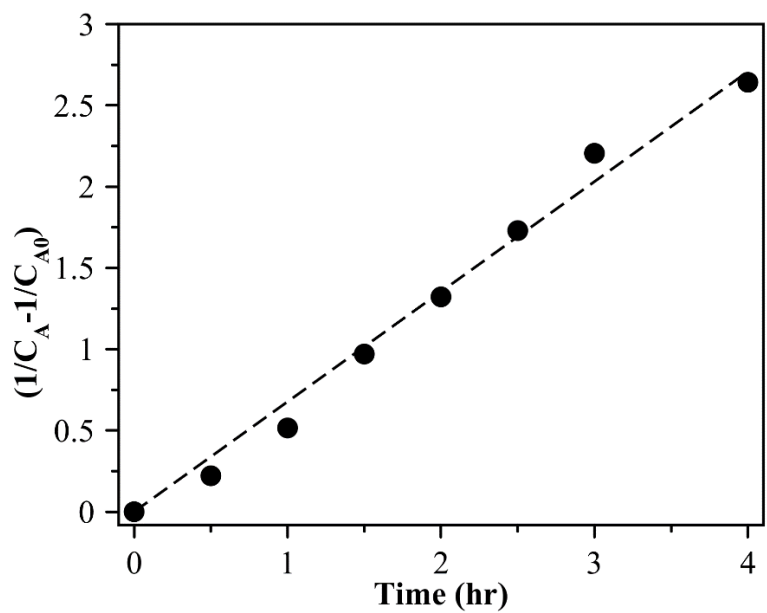

**Figure S4 (w)** Plot of Empirical fit with second order $(1/C_A-1/C_{A0})$ versus irradiation time for MB DSD in DMF, using quasi-spherical $Cu_2O$ nanoparticles of $145 \pm 41$ nm diameter and corresponding trendline and proportion of the variance.

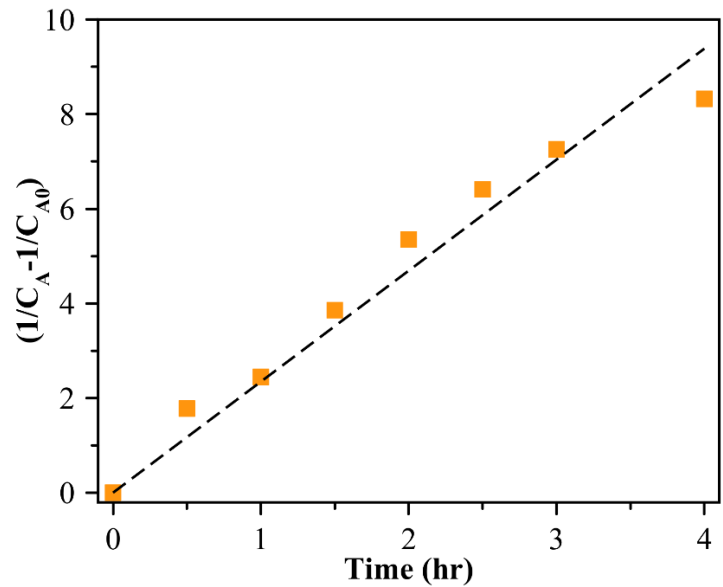

**Figure S4 (x)** Plot of Empirical fit with second order $(1/C_A - 1/C_{A0})$ versus irradiation time for MB DSD in DMF, using $Cu_2O$ nanocubes of $165 \pm 26$ nm edge length and corresponding trendline and proportion of the variance.

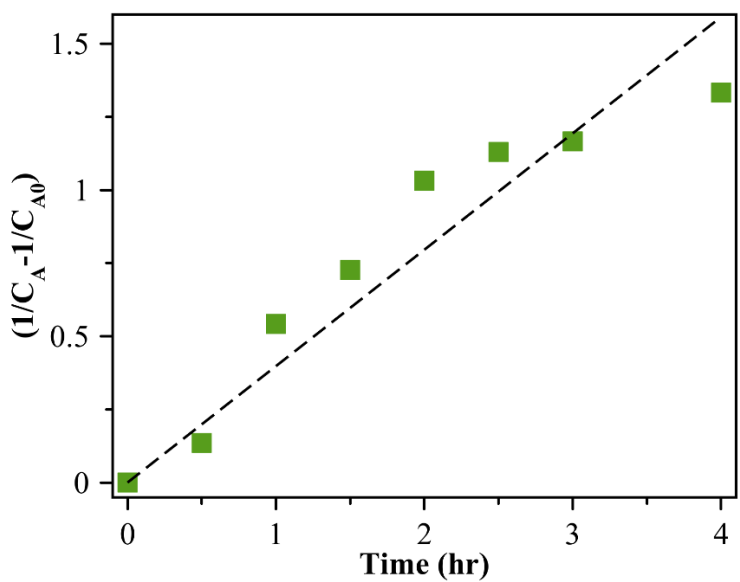

**Figure S4 (y)** Plot of Empirical fit with second order $(1/C_A - 1/C_{A0})$ versus irradiation time for MB DSD in DMF, using $Cu_2O$ nanocubes of $325 \pm 37$ nm edge length and corresponding trendline and proportion of the variance.

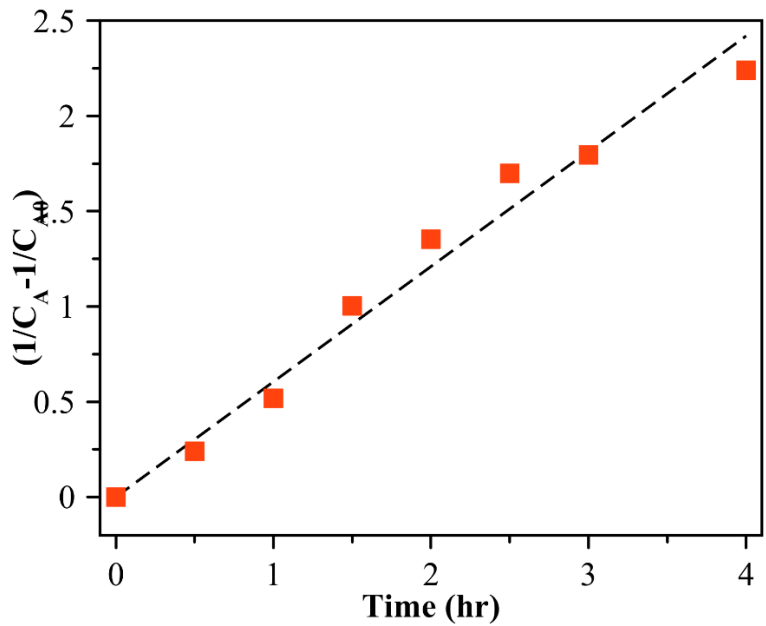

**Figure S4 (z)** Plot of Empirical fit with second order ($1/C_A$-$1/C_{A0}$) versus irradiation time for MB DSD in DMF, using $Cu_2O$ nanocubes of $92 \pm 13$ nm edge length and corresponding trendline and proportion of the variance.

**Table S1**. Fitted second-order rate constant values obtained from the MB dye-sensitization followed by degradation using $Cu_2O$ nanospheres and nanocubes of different sizes.

| $Cu_2O$ Nanostructures | Apparent Rate Constant, $kC_{A0}$ (hr$^{-1}$) |
|---|---|
| Spheres ($37\pm6$ nm) | $0.2548\pm0.0227$ |
| Spheres ($145\pm41$ nm) | $0.637\pm0.0761$ |
| Cubes ($92\pm13$ nm) | $0.676\pm0.083$ |
| Cubes ($165\pm26$ nm) | $2.293\pm0.598$ |
| Cubes ($325\pm37$ nm) | $0.394\pm0.0253$ |

**References Cited in Supporting Information**